\documentclass{emulateapj}
\shorttitle{Star-forming galaxies in the Redshift Desert}
\shortauthors{Steidel et al.}
\slugcomment{Received 2003 November 6; Accepted 2003 December 9}
\newcommand{\ha}{H$\alpha$}
\newcommand{\Ha}{\,{\rm H\alpha}}
\newcommand{\lya}{Ly$\alpha$}

\newcommand{\kms}{\,km~s$^{-1}$}      
\newcommand{\minpoint}{\mbox{$'\mskip-4.7mu.\mskip0.8mu$}}

\newcommand{\secpoint}{\mbox{$''\mskip-7.6mu.\,$}}

\newcommand{\et}{{\rm et al.}~}
\newcommand{\msun}{\,{\rm M_\odot}}
\def\ltsima{$\; \buildrel < \over \sim \;$}
\def\simlt{\lower.5ex\hbox{\ltsima}}
\def\gtsima{$\; \buildrel > \over \sim \;$}
\def\simgt{\lower.5ex\hbox{\gtsima}}
\def\arcs{$''~$}
\def\arcm{$'~$}

\begin{document}
\title{A SURVEY OF STAR-FORMING GALAXIES IN THE 
$1.4 \simlt Z \simlt 2.5$ `REDSHIFT DESERT': OVERVIEW \altaffilmark{1} }
\author{\sc Charles C. Steidel \& Alice E. Shapley\altaffilmark{2}}
\affil{California Institute of Technology, MS 105--24, Pasadena, CA 91125}
\author{\sc Max Pettini}
\affil{Institute of Astronomy, Madingley Road, Cambridge CB3 OHA, UK}
\author{\sc Kurt L. Adelberger}
\affil{Carnegie Observatories, 813 Santa Barbara Street, Pasadena, CA 91101}
\author{\sc Dawn K. Erb, Naveen A. Reddy, \& Matthew P. Hunt} 
\affil{California Institute of Technology, MS 105--24, Pasadena, CA 91125}


\altaffiltext{1}{Based, in part, on data obtained at the 
W.M. Keck Observatory, which 
is operated as a scientific partnership among the California Institute of 
Technology, the
University of California, and NASA, and was made possible by the generous 
financial
support of the W.M. Keck Foundation.
} 
\altaffiltext{2}{Present address: Department of Astronomy, University of 
California, Berkeley, CA 94720}
\begin{abstract}
The redshift interval $1.4 \simlt z \simlt 2.5$
has been described by some as the `redshift desert' because of
historical difficulties in spectroscopically identifying galaxies in that range.  
In fact, galaxies can be found in large numbers with standard broad-band color
selection techniques coupled to follow-up spectroscopy with
UV and blue-sensitive spectrographs. In this paper we present
the first results of a large-scale survey of such objects,
carried out with the blue channel of the LRIS spectrograph (LRIS-B) on the 
Keck~I telescope. We introduce two samples of star forming galaxies,
`BX' galaxies at $\langle z \rangle = 2.20 \pm 0.32$
and `BM' galaxies at  $\langle z \rangle = 1.70 \pm 0.34$.
In seven survey fields we have spectroscopically confirmed 749 of 
the former and 114 of the latter. 
Interlopers (defined as objects at $z < 1$)
account for less than 10\% of the photometric 
candidates, and the fraction of faint AGN is 
$\sim 3$\% in the combined BX/BM sample.
Deep near-IR photometry of a subset of the BX sample 
indicates that, compared to a sample of similarly
UV-selected galaxies at $z \sim 3$, the $z \sim 2$ galaxies
are on average significantly redder in 
(${\cal R} - K_s$), indicating longer star formation histories,
increased reddening by dust, or both. Using near-IR H$\alpha$ spectra of a subset
of BX/BM galaxies to define the galaxies' systemic redshifts, 
we show that the galactic-scale
winds which are a feature of star-forming
galaxies at $z \sim 3$ are also common at later
epochs and have similar bulk outflow speeds of 200-300\,km~s$^{-1}$.
We illustrate by example the information 
which can be deduced on the stellar populations, 
metallicities, and kinematics of ``redshift desert'' galaxies
from easily accessible rest-frame far-UV and rest-frame optical spectra. 
Far from being hostile to observations, the universe at $z \sim 2$
is uniquely suited to providing information on the astrophysics
of star-forming galaxies and the intergalactic medium, and the relationship between 
the two.
\end{abstract}

\section{INTRODUCTION}

A number of different observations point to the redshift range $1 \simlt z 
\simlt 2.5$ as a particularly
important epoch in the history of star formation, accretion onto massive 
black holes, and galaxy assembly.  
Recent successes in identifying the luminous but heavily obscured galaxies 
selected at sub-mm and radio wavelengths have shown that 
they are mostly at redshifts $z \sim 2.4 \pm 0.4$ (Chapman \et 2003);
this is also the epoch when the number density of luminous quasi-stellar
objects (QSOs) peaked (e.g., Di Matteo et al. 2003 and references therein).
The evolution of the ultraviolet (UV) luminosity density of the universe is 
now mapped out with reasonable precision over the redshift 
range $z \sim 0-6$ (e.g., Madau \et 1996; Connolly \et 1997; Steidel \et 1999; 
Giavalisco et al. 2003), but there remains a glaring gap
between $z \sim 1.5$ and $z \sim 2.5$, an epoch when 
much of today's stellar mass was assembled and heavy elements were produced
(e.g., Dickinson \et 2003; Fontana \et 2003; Rudnick et al. 2003). 
The reason for this gap is that such redshifts, while only `modest'
by current standards in distant galaxy hunting,
have remained challenging to direct observation from the ground.
As explained below, this has only to do with accidental incompatibilities
between technology,  the atmospheric windows available for sensitive observations
from the ground, and the spectral features that enable redshift measurement, 
and not with any intrinsic changes in the
galaxy populations. The fact that the $z=1.5 - 2.5$ universe is to
a large extent still {\it terra incognita} provides
exciting opportunities for new observational techniques 
to make substantial headway.

Spectroscopy of distant galaxies has advanced significantly since the advent 
of 8-10\,m class
telescopes located at the best terrestrial sites and equipped with state of 
the art
spectrographs achieving very high efficiency throughout the optical range. 
Wavelengths between 4000 and 9000\,\AA\ in particular benefit from the
combination of low night-sky background, high atmospheric transmissivity, 
and high CCD quantum efficiency which all result
in greater sensitivities (in flux density units) than those
achievable at any other wavelength observable from the ground. 
For this reason, galaxy surveys 
have traditionally targeted strong spectral features 
that fall in this range of high sensitivity, 
particularly nebular lines from \ion{H}{2} regions 
(e.g., [\ion{O}{2}]~$\lambda 3727$;
[\ion{O}{3}]~$\lambda\lambda 4959, 5007$; H$\beta$ and H$\alpha$), 
and the region near 4000\,\AA\ in the spectra of early type galaxies. 
These features are used both in the measurement of galaxy redshifts
and in the determination of basic astrophysical properties, 
such as star-formation rates,  
reddening, chemical abundances, velocity dispersions, and age.

As we move from the local universe to higher redshifts, 
the most straightforward strategy has been
to simply follow these same spectral lines 
to longer wavelengths; for redshift measurements,
[\ion{O}{2}]~$\lambda 3727$  is accessible to optical spectrographs
up to $z \simeq 1.4$. 
Considerable success in charting galaxies up to these redshifts
has been achieved recently with  red-optimized\footnote{Red optimization 
involves several aspects: high spectral resolution, 
low detector fringing and/or spectrograph stability, as well as high detector
quantum efficiency in the red.} spectrographs (e.g.,  DEEP2, Coil \et 2003) 
that have been designed specifically 
to minimize the effects of OH emission from the night sky
at wavelengths longer than $\sim 7300$\,\AA, or $z \simgt 1$ for
the detection of [\ion{O}{2}]~$\lambda 3727$.  

The optical domain has also proved ideal for surveying galaxies at redshifts
$z\simgt 2.5$  where the rest-frame far-UV spectral region, with its wealth of
spectral information, crosses over the `horizon' imposed
by the earth's atmosphere (Steidel \et 1999, 2003; Shapley et al. 2003). 
Between $z \sim 1.4$ and $z \sim 2.5$, however, lies an interval of
redshift for which no strong spectral lines fall 
in the $4300-9000$\,\AA\ range
where most of the spectrographs on large 
telescopes are optimized---hence, the `redshift desert'.
There are two obvious approaches to finding galaxies 
at these redshifts:
one is to extend multi-object spectroscopy into the near-infrared (near-IR)
and target the familiar optical emission lines from \ion{H}{2} regions;
the other is to exploit the rest-frame far-UV 
spectral features by observing at near-UV and blue 
wavelengths ($\sim 3100-4500$\,\AA).

Both strategies are observationally challenging. 
In the first case,
the sky background---both in the OH emission features 
and in the continuum---becomes progressively brighter
at longer wavelengths,
while the falling efficiency of silicon
detectors requires the use of a different detector 
technology---one that lags significantly behind
CCDs in terms of performance, multiplexing, and 
areal coverage. 
Moreover, the increasing importance
of thermal noise in the IR necessitates
more complex cryogenic instruments with 
cooled focal planes and optics. 
Nevertheless, multi-object cryogenic near-IR 
spectrographs are now being built, or planned, for 
several 8-10\,m telescope facilities.
At the other end of the scale, the ground-based 
near-UV spectral region has its own difficulties. 
High near-UV transmission and/or reflectivity requires
compromises that can exacerbate broad-band optical performance. 
Different optical glasses and coatings
are required; it is extremely difficult to achieve 
good optical performance simultaneously at
3300\,\AA\ and 7500\,\AA\ with refractive optics, for example, 
and until recently good UV reflectivity has required
the use of Al mirror coatings which do not perform well,
compared to other readily available materials,
at wavelengths longer than 4500\,\AA. 
In addition, in selecting CCDs one must generally choose
between those that have good UV response and 
those with good quantum efficiency 
and low fringing amplitude in the red and near-IR. 
All these reasons explain why the wavelength region below
4000\,\AA\ has so far been largely neglected in the design 
of optical faint-object spectrographs.

From the point of view of the astrophysical information
they convey, both the rest-frame far-UV and optical
regions are important and are in fact largely complementary.
In the rest-frame optical, nebular emission lines yield information
on the chemical abundances and kinematics of the ionized
gas (e.g., Teplitz et al. 2000; Pettini et al. 2001; Erb \et 2003;
Lemoine-Busserolle et al. 2003).
In the far-UV, on the other hand, 
large number of stellar and (especially) 
interstellar absorption lines are accessible.
The latter have provided information on the
kinematics and chemistry of outflowing gas 
that may have very significant implications for the
galaxy formation process 
(see, e.g., Pettini \et 2002a; Shapley \et 2003, Adelberger \et 2003). 
Stellar features, while weaker and therefore more difficult to detect, can 
be used to place constraints on the initial mass function (IMF) 
and metallicity of massive stars 
(e.g., Pettini \et 2000, 2002b; Leitherer et al. 2001).  

When it comes to identifying galaxies at $z = 1.4 - 2.5$ in large
numbers, however, there are clear differences between
the near-UV and near-IR (in the observed frame) domains.
Of course, some types of objects,
such as those most heavily reddened by dust or having no 
current star formation, may be accessible only 
in the near-IR. However, the fact remains that,
with no moonlight, the night sky background in the
blue and near-UV is nearly featureless, is $\sim 3$ (AB) magnitudes fainter 
(per square arc second in the continuum) than at 9000 \AA, and is
more than 5 magnitudes fainter than in the {\it H} (1.65\,$\mu$m)  
and {\it K} (2.2\,$\mu$m) bands {\it even between the OH sky 
emission lines}.
Since the spectral energy distribution of a typical star forming galaxy 
is relatively flat from the rest-frame UV to the optical, the advantage of 
near-UV spectroscopy is obvious, 
provided high spectral throughput can be achieved.
This is most effectively accomplished with double-beam spectrographs
where the spectral throughput can be optimized over the whole range
$0.3-1.0 \mu$m.  

Since the commissioning of the blue channel of the Low Dispersion
Imaging Spectrograph (LRIS-B) on the Keck~I telescope, described
in an appendix, we have been conducting a survey for star-forming
galaxies at redshifts $z \simeq 1.5 - 2.5$. The development of
the photometric criteria used to select candidates is
described in detail in Adelberger \et (2004). In this paper we present
the first results of the survey, highlighting the efficiency
of UV selection for bridging this important redshift gap
in our knowledge of galaxy evolution, and and overview of the science
now possible for galaxies in this redshift range. 
In \S2 we briefly describe
our photometric selection of candidates. 
The spectroscopy is described in \S3 where we present the first
results of the survey, such as the success rate of the photometric
selection and the redshift distribution of confirmed candidates. 
An appendix describes the most important aspects of the LRIS-B instrument,
which has been crucial to the success of the survey. 
\S4 illustrates some of the astrophysical information 
conveyed by the spectra; we consider in particular the
IMF and metallicity of the young stellar populations and 
the kinematics of the interstellar medium in these galaxies.
\S5 deals with the optical-IR colors of the galaxies.
Finally, in \S6  we summarize the main findings from
this initial stage of our survey. We assume a cosmology with
$\Omega_m=0.3$, $\Omega_{\Lambda}=0.7$, and $h=0.7$ throughout.

\section{PHOTOMETRIC AND SPECTROSCOPIC TARGET SELECTION}

\subsection{Color Selection}

Our searches for galaxies at $z = 1.4 - 2.5$ 
are based on deep images in the $U_n$, $G$, and ${\cal R}$ passbands
of similar quality (in terms of depth and seeing)
as those used in our published survey for Lyman Break
Galaxies (LBGs) at $z \simeq 3$ (Steidel \et 2003).
The fields observed are listed in Table~1;
with the exception of GOODS/HDF-N and
Westphal, the fields are distinct from those used in the $z \sim 3$ LBG survey and were selected 
primarily because they include one or more relatively bright background QSOs
suitable for studying the cross-correlation of
galaxies with \ion{H}{1} and metals in the intergalactic
medium (IGM). The imaging data were obtained at four
telescopes (Palomar 5.1\,m, WHT 4.2\,m, KPNO 4\,m, and
Keck I 10\,m) mostly between 2000 August and 2003 April.
We made use of the deep $U$ band image in the GOODS-N field 
obtained by the GOODS team (Giavalisco \& Dickinson 2003), which was calibrated onto
our own photometric system using observations 
through the $U_n$ filter presented in Steidel \et (2003). 
The image reductions and photometry were performed 
following the procedures described in Steidel \et (2003). 

\begin{deluxetable*}{lcccl}[!tbp]
\tablewidth{0pc}
\tablecaption{Survey Fields}
\tablehead{
\colhead{Field Name} & \colhead{$\alpha$(J2000)} & \colhead{$\delta$(J2000)} & \colhead{Field Size}  
& \colhead{Telescope/Date\tablenotemark{a}} 
} 
\startdata
GOODS-N & 12:36:51 & $+$62:13:14 & 10\minpoint6$\times$14\minpoint6 & KPNO/Apr 02, Keck~I/Apr 03 \\ 
Q1307   & 13:07:45 & $+$29:12:51 & 16\minpoint2$\times$15\minpoint9  & WHT/May 01 \\ 
Westphal & 14:17:43 & $+$52:28:49 & 15\minpoint0$\times$15\minpoint0 & KPNO/May 96, P200/May 02 \\
Q1623    & 16:25:45 & $+$26:47:23 & 12\minpoint5$\times$23\minpoint2 & P200/Aug 00 \\
Q1700    & 17:01:01 & $+$64:11:58 & 15\minpoint3$\times$15\minpoint3 & WHT/May 01 \\
Q2343    & 23:46:05 & $+$12:49:12 & 23\minpoint8$\times$11\minpoint9 & P200/Aug 01 \\
Q2346    & 23:48:23 & $+$00:27:15 & 16\minpoint4$\times$17\minpoint0 & WHT/Aug 01 \\
\enddata
\tablenotetext{a}{Telescope and instrument used to obtain the deep imaging data: 
KPNO\,$=$\,Kitt Peak 4\,m Mayall telescope, with PFCCD (1996) and MOSAIC (2002);
WHT\,$=$\,William Herschel 4.2\,m telescope with prime focus imager; 
P200\,$=$\,Palomar 5.1\,m Hale telescope with Large Format Camera;
Keck~I\,$=$\,Keck~I 10\,m telescope LRIS in imaging mode.}
\end{deluxetable*}

The rationale and method for selecting $z \sim 2$ galaxies using only 
their optical broad-band colors are described in detail by Adelberger \et 
(2004). 
The selection criteria are aimed at identifying galaxies with approximately
the same range of intrinsic properties, particularly UV luminosity and 
reddening
by dust, as the well-studied $z \sim 3$ Lyman break galaxies.
After some `fine-tuning' based on the initial results 
of early spectroscopic follow-ups,
we converged on two sets of color selection criteria designed
respectively to select galaxies in the redshift ranges 
$2.0 \simlt z \simlt 2.5$---we call these `BX' objects, 
and $1.5 \simlt z \simlt 2.0$---the `BM' objects. 
The criteria for BX objects (e.g., Q1700-BX691) are
\footnote{Note 
that the BX color cuts given in equation 1 are slightly
different from the preliminary values published
in Erb et al. (2003).}: 
\begin{eqnarray}
G-{\cal R} &\geq& -0.2\nonumber\\
U_n-G      &\geq& G-{\cal R}+0.2\nonumber\\
G-{\cal R} &\leq &0.2(U_n-G)+0.4\nonumber\\
U_n-G      &\leq& G-{\cal R}+1.0
\end{eqnarray}
and for the BM objects (e.g., Q1307-BM1163), 
\begin{eqnarray}
G-{\cal R} &\geq& -0.2\nonumber\\
U_n-G      &\geq& G-{\cal R}-0.1\nonumber\\
G-{\cal R} &\leq &0.2(U_n-G)+0.4\nonumber\\
U_n-G      &\leq& G-{\cal R}+0.2
\end{eqnarray}
The methods for establishing these selection criteria, 
and their estimated level of completeness, 
are discussed in Adelberger \et (2004). 
Figure 1 shows where the two color cuts are located in the 
$(U_n-G)$ vs. $(G-\cal R)$ plane.
The resulting sample of galaxies is very similar to
our existing sample of $z \sim 3$ LBGs in terms
of star formation rate inferred
from their UV luminosities (uncorrected for extinction):
the BX+BM spectroscopic sample has SFR\,$=3-60\,M_{\sun}$~yr$^{-1}$,
with a median of $9.9\,M_{\sun}$~yr$^{-1}$,
while the spectroscopic $z \sim 3$ LBGs have 
SFR\,$=5.5-66\,M_{\sun}$~yr$^{-1}$ 
with a median of $10.3\,M_{\sun}$~yr$^{-1}$
(estimated from the apparent $G$ and ${\cal R}$ magnitudes
respectively, using the conversion from 1500\AA\ luminosity 
advocated by Kennicutt 1998.)

\begin{figure}[htb]
\centerline{\epsfxsize=9cm\epsffile{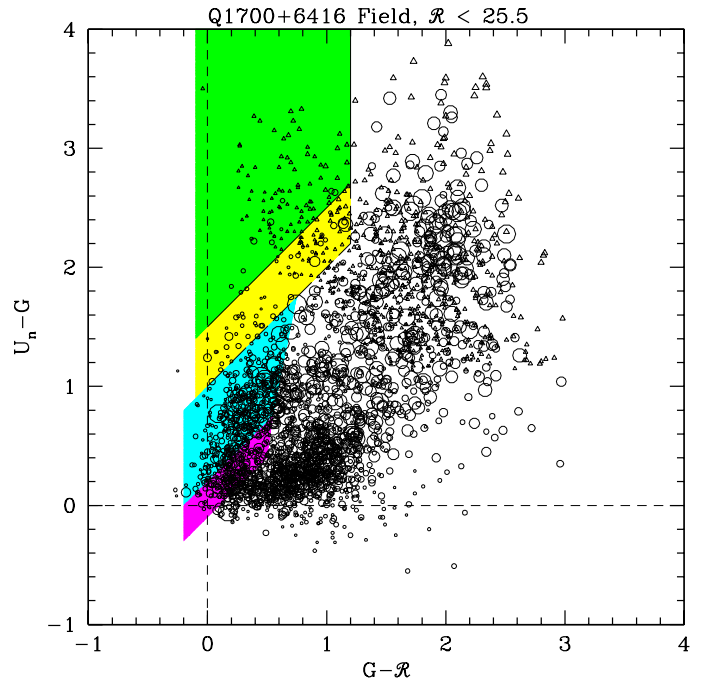}}
\figcaption[f1.eps]{Two-color diagram from one of the survey fields, to illustrate the
new selection criteria discussed in this paper (see also Adelberger \et 2004). 
The size of the symbols scales with object brightness, and
triangles are objects for which only limits were obtained on the $U_n-G$ color.
The cyan and magenta-shaded regions are the BX and BM selection windows,
designed to select galaxies at $z \simeq 2-2.5$ and $z \simeq 1.5-2.0$, respectively. The
green and yellow shaded regions are the $z \sim 3$ LBG color section windows used in the
survey by Steidel \et (2003).
In this field there are 1831 BX and 1085 BM candidates;
together they account for $\sim 25$\% of the 11547 objects brighter
than ${\cal R} = 25.5$. For clarity only 
one in five objects is shown in the plot.
}
 
\end{figure}

The average surface density of photometric candidates 
satisfying the BX criteria is 5.2\,arcmin$^{-2}$ 
to ${\cal R}=25.5$\footnote{Corrected for 9\% contamination by interlopers, 
as discussed in \S4.1 below.}, 
while the corresponding average surface density of BM objects
is 3.8\,arcmin$^{-2}$; together, they comprise $\sim 25$\% of the ${\cal R}$ 
band counts to this apparent magnitude limit, 
and exceed the $z \sim 3$ LBG surface density by a factor of
more than five. 
This is not surprising given the larger redshift interval 
($1.5 \simlt z \simlt 2.5$), and the lower
intrinsic luminosities reached in the BX and BM samples 
compared to the $z \sim 3$ LBGs.\footnote{${\cal R}=25.5$ corresponds to 
galaxies that are 0.6 magnitudes less luminous at $z=2.2$, and 
1.1 mags less luminous at $z=1.7$, than $z \sim 3$ LBGs of the 
same apparent magnitude.}

\subsection{Spectroscopic Follow-up}
At this stage in the survey we have followed-up spectroscopically
primarily the BX candidates, for the purpose of extending to lower
redshifts our study of the IGM-galaxy connection begun in
Adelberger et al. (2003). As discussed in \S3.1,
the redshift distribution of these galaxies has  
$\langle z \rangle = 2.20\pm0.32$.
Only very recently have we begun observing significant numbers of 
BM galaxies targeting the range $1.4 \simlt z \simlt 2.0$,
considered by some to be perhaps the most challenging
range of redshifts for confirmation with optical spectroscopy.
While the current statistics for such galaxies
are not as extensive as those of the BX sub-sample
(we have observed only 187 candidates to date), 
the results obtained so far already show that the color selection 
criteria work as expected; the redshift distribution of 
confirmed BM objects 
is $\langle z \rangle = 1.70 \pm 0.34$. 

What is clear is that the higher surface density of 
BX/BM photometric candidates, 
coupled with the high rate of spectroscopic confirmation achieved
with the UV-optimized LRIS-B spectrograph,
allows a large sample of galaxies in the `redshift desert'
to be assembled relatively easily and efficiently.
To date our survey includes 692 galaxies with confirmed
spectroscopic redshifts between $z = 1.4$ and 2.5\,.
We defer to a future paper the relatively complex analysis necessary 
to turn the photometric and spectroscopic
results into a far-UV luminosity function of $z \sim 2$ galaxies.

We now turn to the spectroscopic results; relevant information
on the performance and specifications of the LRIS-B instrument
are summarized in the Appendix.

\section{SPECTROSCOPIC OBSERVATIONS AND RESULTS}

\subsection{Optical Spectroscopy}

For the redshift range of interest here, $1.4 \simlt z \simlt 2.5$,
the rest frame far-UV region between Ly$\alpha$ and 
\ion{C}{4}~$\lambda\lambda 1548,1550$, with its rich complement of
stellar and interstellar lines, is redshifted 
between the atmospheric cut-off near 3100\,\AA\ and 
5400\,\AA. This wavelength interval was therefore 
the primary target of our spectroscopic survey. 
The spectra were obtained with slightly different 
instrumental configurations over the three year period 
of the survey to date.
In the blue channel of LRIS we used  
the 400 groove~mm$^{-1}$ grism blazed at 3400\,\AA\ throughout,
since this is the grism which provides the highest 
throughput between 3100 and 4000\,\AA\ (see Appendix and associated
figures).
Initially we observed with a dichroic that divides 
the incoming beam at 5600\,\AA, and with a 600 groove~mm$^{-1}$ grating 
blazed at 7500\,\AA\ on the red side; this set-up
generally gives complete wavelength coverage
over the entire 3100--8000\,\AA\ range for most slits.
In 2002, once the blue channel detector was upgraded to the
larger Marconi CCD mosaic, 
we used the blue side only with a mirror in place of the dichroic;
this configuration covers from the atmospheric cut-off to about
6500\,\AA\ (the exact red limit of the spectra depends on the
location of a given slit within the field of view)
with a dispersion of 1.07\,\AA\ per 15\,$\mu$m pixel.
More recently we have reverted to double-channel mode, 
but with the 6800\,\AA\ dichroic and 
the 400 groove~mm$^{-1}$ grating blazed at 8500\,\AA\ 
which gives additional spectral coverage from 6800\,\AA\ to 9500\,\AA\ for
most slits. This latest setup is particularly useful when observing 
BM candidates, because the red side allows one to check for the presence
of [\ion{O}{2}]~$\lambda3727$ emission at $z \sim 1 - 1.5$; 
as explained below, galaxies in this
redshift range turn out to be the main source of contamination 
of the BM sample.

The Keck~I f/15 Cassegrain focus does not yet have an 
atmospheric dispersion corrector, so we took
special care to avoid significant light losses due to 
atmospheric dispersion. We designed
slit masks to be used at a given time of night, 
with a position angle that would place the slits within 20 degrees
of the parallactic angle at any time during the observations. 
At the telescope, we performed the final mask alignment in the blue,
since this is the region of most interest for our purposes. 
With the 1.2\,\arcsec\ slits used in all the masks, 
and the typical image quality of $\sim 0.8$\,\arcsec\ at the detector, 
the spectral resolution in the blue was $\simeq 5$\,\AA\ FWHM,
sampled with $\sim 4.5$ pixels.  
Wavelength calibration was achieved by comparison with
the spectra of internal Cd, Zn, Ne, and Hg lamps,
as well as by reference to night sky emission lines.

The choice of BX and BM candidates assigned to each mask
was based on an algorithm that gives largest weights 
to objects in the apparent magnitude
range ${\cal R}=22.5-24.5$ and somewhat lower weights 
to brighter and fainter objects.
\footnote{Objects lying 
within 1--2\,\arcmin\ of a background QSO
were given additional weight depending on the projected 
distance from the QSO. Such cases are particularly useful for
investigating the galaxy-IGM connection (Adelberger et al. 2003).} 
Each slit mask covered an area on the sky of 8\minpoint0 by 4\minpoint5, 
and contained $30-35$ slits. 
With few exceptions, each mask was observed for 
a total of 5400\,s split into three 1800\,s integrations; 
the telescope was stepped by $1-2$\,\arcsec\ in the slit direction 
between exposures. 
The data were reduced and the spectra identified 
with the procedures described by Steidel \et (2003). 

The total exposure times were deliberately kept short in 
order to maximize the number of galaxies for
which redshifts could be measured. 
Because of this, the quality of the spectra varies considerably
(since the objects range from ${\cal R}=21.7$ to ${\cal R}=25.5$),
but the best spectra are already suitable for more detailed analyses, 
as discussed in \S4 below. The success rate in
spectroscopically identifying candidates varied from mask
to mask, depending on the observing conditions.
On masks obtained in the best conditions, 
a redshift could be measured for more than 90\% of the objects targeted,
whereas the proportion was lower for masks observed in poor
seeing (FWHM\,$\simgt 1$\,arcsec) or through thick cirrus.
Because of this, we believe that spectroscopic failures 
are very likely to have the same redshift distribution
as the successes, rather than being the result of the true 
redshifts falling far from expectations. 
The overall success rate to date in securing
a spectroscopic identification is 69\% for BX candidates 
and 65\% for BM candidates (see tables 2 and 3). 

\begin{deluxetable*}{lccccc}[!btp]
\tablewidth{0pc}
\tablecaption{BX Spectroscopic Sample Completeness and Contamination}
\tablehead{
\colhead{${\cal R}$ Mag} & \colhead{Attempted} & \colhead{Identified} & \colhead{\% Identified}  
& \colhead{Interlopers\tablenotemark{a}} & \colhead{\% Interlopers\tablenotemark{b}} 
} 
\startdata
$19.0-22.0$\tablenotemark{c} & ~~52 & ~50 & ~96.2 & ~42 & 84.6 \\
$22.0-22.5$ & ~~12 & ~12 & 100.0 & ~11 & 91.7 \\
$22.5-23.0$ & ~~32 & ~30 & ~93.8 & ~21 & 70.0 \\
$23.0-23.5$ & ~118 & 106 & ~89.8 & ~32 & 30.2 \\
$23.5-24.0$ & ~223 & 177 & ~79.3 & ~25 & 14.0 \\
$24.0-24.5$ & ~363 & 248 & ~68.3 & ~12 & ~4.8 \\
$24.5-25.0$ & ~275 & 175 & ~63.6 & ~~9 & ~5.1 \\
$25.0-25.5$ & ~231 & 105 & ~45.7 & ~~2 & ~1.9 \\ 
Total       & 1309 & 903 & ~69.1 & 154 & 17.1 \\
\enddata
\tablenotetext{a}{Number of objects with $z < 1$.}
\tablenotetext {b}{Fraction of identified objects with $z < 1$.}
\tablenotetext{c}{7 of the 8 objects with ${\cal R}<22$ and $z > 1$ are QSOs.}
\end{deluxetable*}

\begin{figure}[htb]
\centerline{\epsfxsize=9cm\epsffile{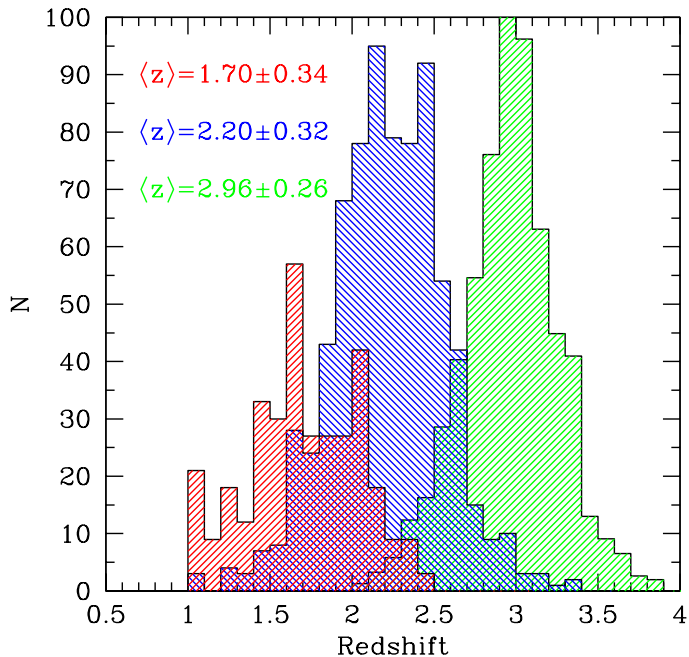}}
\figcaption[f2.eps]{Redshift distributions of spectroscopically confirmed
galaxies from various color-selected samples. The green
histogram is for $z \sim 3$ LBGs from Steidel \et (2003) (scaled by 0.7); 
the blue histogram shows the redshifts of our current BX sample
(749 galaxies), while the red histogram is the smaller BM sample
of 114 galaxies (scaled up by a factor of three for clarity).
The total number of new spectroscopic redshifts 
in the range $1.4 \le z \le 2.5$, including both
BX and BM samples, is 694, 
with 244 of those in the range $1.4 \le z \le 2.0$.}
\end{figure}

The current redshift histograms for the BX and BM samples 
are shown in Fig. 2, together with the histogram for the 
completed $z\sim 3$ LBG survey of Steidel et al. (2003).
The total number of new spectroscopic redshifts 
in the range $1.4 \le z \le 2.5$, including both
BX and BM samples, is 692, 
with 244 of those in the range $1.4 \le z \le 2.0$.

\begin{deluxetable*}{lccccc}[!tbp]
\tablewidth{0pc}
\tablecaption{BM Spectroscopic Sample Completeness and Contamination}
\tablehead{
\colhead{${\cal R}$ Mag} & \colhead{Attempted} & \colhead{Identified} & \colhead{\% Identified}  
& \colhead{Interlopers\tablenotemark{a}} & \colhead{\% Interlopers\tablenotemark{b}} 
} 
\startdata
$19.0-22.0$\tablenotemark{c} & ~~7 & ~7 & 100.0 & 0 & ~0.0 \\
$22.0-22.5$ & ~~1 & ~1 & 100.0 & 0 & ~0.0 \\
$22.5-23.0$ & ~~2 & ~2 & 100.0 & 0 & ~0.0 \\
$23.0-23.5$ & ~21 & 17 & ~81.0 & 1 & ~5.8 \\
$23.5-24.0$ & ~38 & 29 & ~76.3 & 3 & 10.3 \\
$24.0-24.5$ & ~67 & 40 & ~59.7 & 3 & ~7.5 \\
$24.5-25.0$ & ~30 & 17 & ~56.7 & 0 & ~0.0 \\
$25.0-25.5$ & ~21 & ~~8 & ~38.1 & 0 & ~0.0 \\ 
Total       & 187 & 121 & ~64.7 & 7 & ~5.8 \\
\enddata
\tablenotetext{a}{Number of objects with $z < 1$.}
\tablenotetext {b}{Fraction of identified objects with $z < 1$.}
\tablenotetext{c}{6 of the 7 objects with ${\cal R}<22$ and $z > 1$ are QSOs.}
\end{deluxetable*}

From the results of the survey so far we have established that
star forming galaxies at $\langle z \rangle= 0.17 \pm 0.09$
are the main source of contamination in the BX sample
(adopting the somewhat arbitrary definition of an interloper 
as any object at $z < 1$).
These are star forming dwarf galaxies 
whose Balmer break mimics the Lyman $\alpha$ forest decrement of
galaxies in the $z=2-2.5$ range. As can be seen from Table 2,
they are dominant at the bright end of the distribution of
apparent magnitudes but become negligible at faint magnitudes. 
Confining the spectroscopic follow-up to objects 
with ${\cal R} >23.5$ would reduce the contamination to 5\%, 
but at the expense of overlooking the intrinsically brightest 
galaxies at $z > 1.5$ which are the most suitable for subsequent 
detailed spectroscopic studies.
For this reason we have included many of these bright objects
in the masks observed up to now.
Based on the spectroscopic results, 
we estimate that the overall contamination of the BX photometric sample
to ${\cal R}=25.5$ by low-redshift interlopers is $\sim 9$\% 
of which 3\% are stars and 6\% are low redshift star forming galaxies. 
However, because we did not sample the photometric candidates evenly, 
but favored brighter objects, the interloper contamination of the 
BX spectroscopic sample to date is $\simeq 17$\%. 
Of the 903 BX objects with measured redshifts, 
749 are at $z > 1$ with $\langle z \rangle = 2.20\pm 0.32$. 

Objects satisfying the BM color criteria, on the other hand, 
do not suffer significant contamination from very low redshift galaxies
or stars. While the targeted redshift range was $1.5 \simlt z \simlt 2.0$, a
significant number of galaxies in the $1.0 \le z \le 1.4$ range are included 
simply because some fraction of such galaxies are found in the same region
of the $U_nG{\cal R}$ color space as the targeted objects (see Table 3);
their proportion
can be reduced if additional color criteria are imposed
(see Adelberger \et 2004). 
Of the 114 BM objects with secure redshifts,
107 have $z > 1$, with an overall redshift distribution
of $\langle z \rangle =1.70 \pm 0.34$; 
20 objects have $1.0 \le z \le 1.4$.

\begin{figure}[htb]
\centerline{\epsfxsize=9cm\epsffile{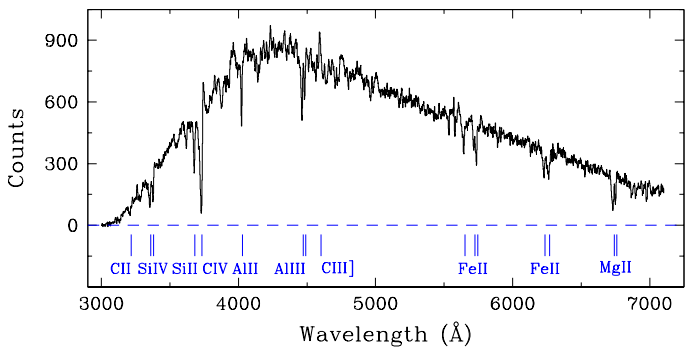}}
\figcaption[f3.eps]{LRIS-B spectrum of Q1307-BM1163, a bright 
(${\cal R} = 21.67$, $G = 21.87$, $U_n = 22.22$)
galaxy at $z = 1.411$. The spectral resolution is $\sim 5$\,\AA\ FWHM.
We have deliberately shown this spectrum
before flux calibration to illustrate how the count rate varies
over the typical wavelength range spanned 
by LRIS-B. Like nearly all of the spectra obtained in the survey, 
this is the sum of three 1800\,s exposures.
The most prominent interstellar absorption and emission lines
used to measure galaxy redshifts between $z = 1.4$ and 2.5
are indicated below the spectrum. 
}
\end{figure}

Example spectra of galaxies in the `redshift desert' are shown
in Figs. 3 and 4; the properties of these example objects,
including their positions, magnitudes, and colors, are summarized in
Table 4. In Fig. 3 we have reproduced the LRIS-B
spectrum of Q1307-BM1163, a bright 
(${\cal R} = 21.67$, $G = 21.87$, $U_n = 22.22$)
galaxy at $z = 1.411$. We have plotted raw counts
vs. observed wavelength specifically to show
the relative count rate as a function of wavelength
from 3000 to 7100\,\AA, the range encompassed by the 
blue channel of LRIS for this object. Useful signal is obtained
all the way down to 3100 \AA; some of the most prominent
interstellar absorption and emission lines
covered in our spectra of BM and BX galaxies,
from \ion{C}{2}~$\lambda 1334$ to \ion{Mg}{2}~$\lambda\lambda 2796,2803$,
are indicated in the figure. 
In \S4 we use this example  to illustrate the information
which can be deduced on the 
stellar populations and the interstellar medium 
of galaxies at $z = 1.4 - 2.5$ from the study of their
rest-frame UV spectra.

\begin{figure}[htb]
\centerline{\epsfxsize=9cm\epsffile{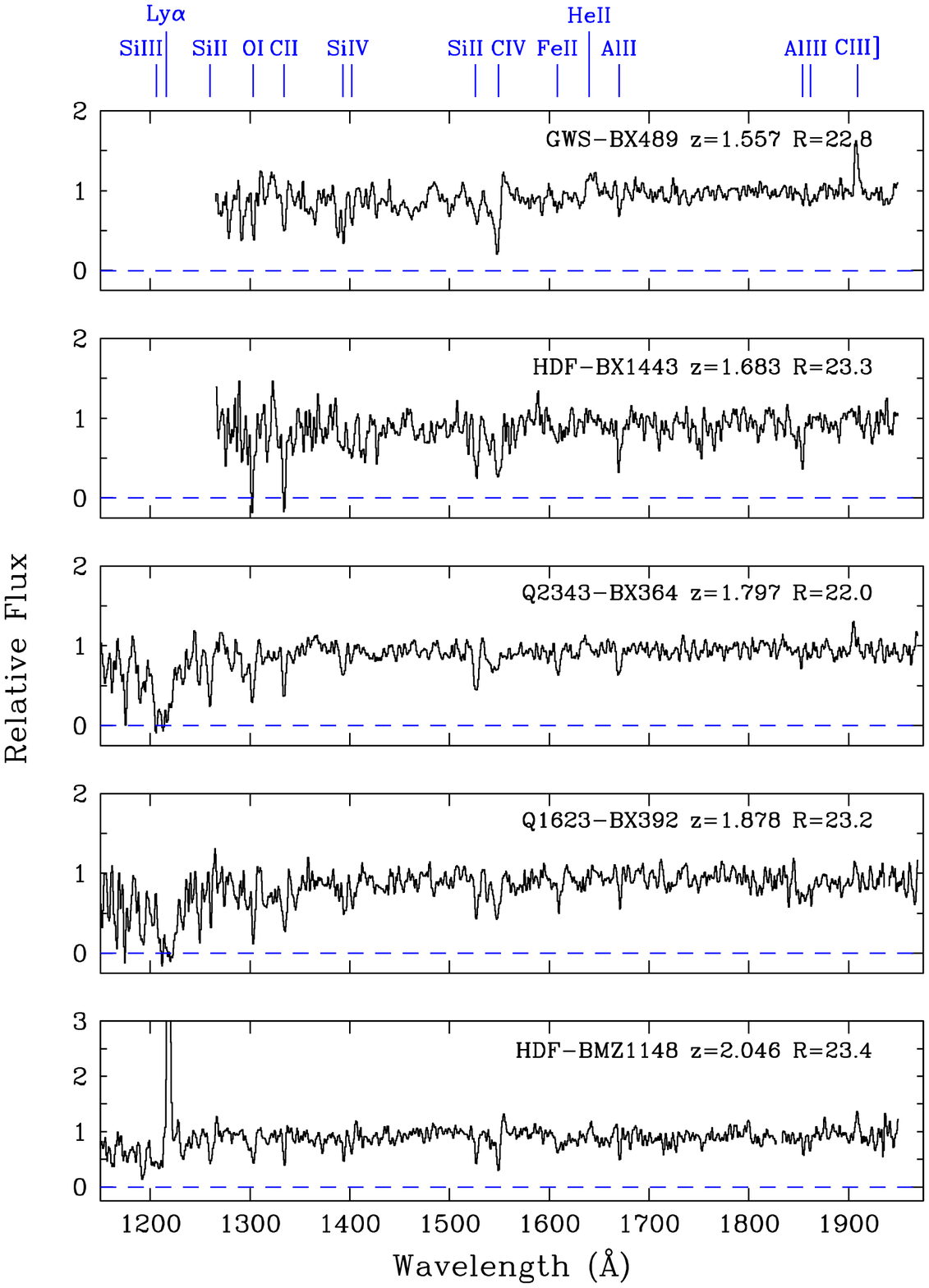}}
\figcaption[f4a.eps]{
Examples of portions of the spectra
of galaxies in the redshift range
$1.5 \simlt z \simlt 2.5$\,.
The spectra have been shifted into the rest frame
and normalized to unity in the continuum.
The strongest spectral features in the spectra of
star forming galaxies at these
redshifts are indicated at the top of each panel.
The galaxies shown here are selected from the brightest $\sim 10$\% of the
spectroscopic sample of 863 BX/BM galaxies obtained to date. Typical galaxies
in the sample are $\sim 1$ mag fainter, and their spectra have S/N lower by a factor of 2--3.
All exposures were 5400\,s.}
\end{figure}

\addtocounter{figure}{-1}

\begin{figure}[htb]
\centerline{\epsfxsize=9cm\epsffile{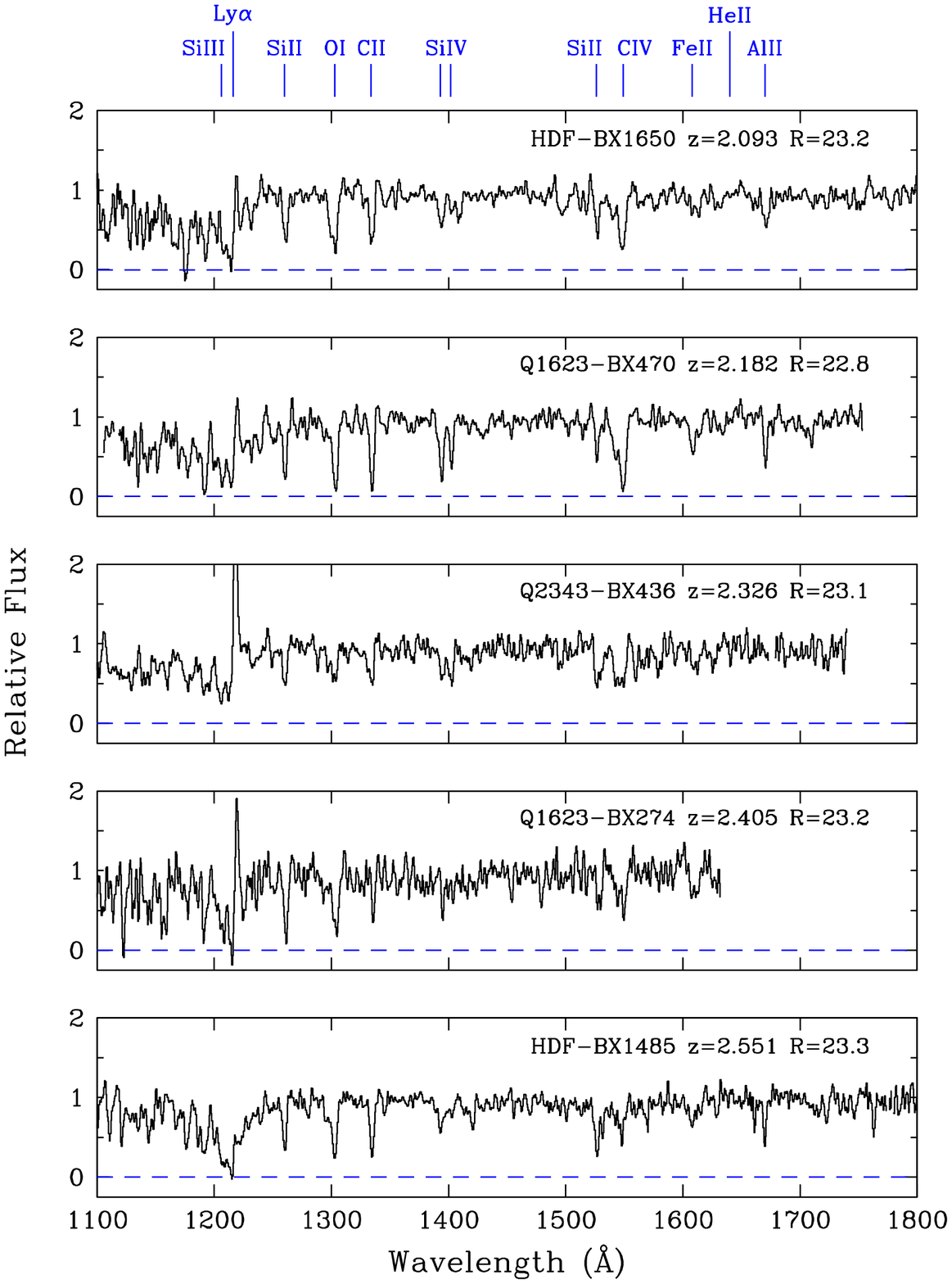}}
\figcaption[f4b.eps]{(Continued).}
\end{figure}

Figure 4 is a montage of ten spectra 
of some of the brighter
BM and BX galaxies chosen to span the range of redshifts
and spectral properties seen among our sample.
Broadly speaking, the far-UV spectra of color-selected 
$z\simeq 1.4-2.5$ galaxies resemble those of Lyman break galaxies
at $z \sim 3$ in terms of continuum slope and spectral lines
seen in emission and absorption. There is some evidence to suggest that 
the Lyman $\alpha$ line appears less frequently in emission
at these redshifts than at $z \simeq 3$:
while 33\% of the $z\sim 3$ LBGs in the survey by Steidel et al. (2003)
have no measurable Lyman $\alpha$ emission, the corresponding
fraction in the current sample of BM and BX galaxies at
$z > 1.7$ (such that Lyman $\alpha$ falls longward of $\sim 3300$\,\AA)
is 57\%. It remains to be established whether this is due to
a subtle selection effect or to real redshift evolution in the escape
fraction of Lyman $\alpha$ photons. We intend
to address this issue in the future with simulations which are
beyond the scope of this paper.

\begin{deluxetable*}{lccccccc}[!btp]
\tablewidth{0pc}
\tablecaption{Example BX/BM Galaxies}
\tablehead{
\colhead{Name} & \colhead{$\alpha$(J2000)} & \colhead{$\delta$(J2000)} & \colhead{${\cal R}$}  
& \colhead{${\rm G}-{\cal R}$} & \colhead{${\rm U_n - G}$} & \colhead{${\rm z_{em}}$\tablenotemark{a}} 
& \colhead{${\rm z_{abs}}$\tablenotemark{b}}
} 
\startdata
Q1307-BM1163 & 13:08:18.06 & $+$29:23:19.2 & 21.66 & 0.20 & 0.35 & \nodata & 1.409 \\
GWS-BX489 & 14:17:20.41 & $+$52:33:18.3 & 22.83 & 0.01 & 0.26 & 1.557 & 1.557 \\
HDF-BX1443 & 12:36:44.87 & $+$62:18:37.9 & 23.33 & 0.31 & 0.57 & \nodata & 1.683 \\
Q2343-BX364 & 23:46:14.97 & $+$12:46:53.9 & 21.96 & 0.31 & 0.71 & \nodata & 1.797 \\
Q1623-BX392 & 16:25:46.56 & $+$26:45:52.3 & 23.23 & 0.11 & 0.54 & \nodata & 1.878 \\
HDF-BMZ1148 & 12:36:46.15 & $+$62:15:51.1 & 23.38 & 0.20 & 0.29 & 2.053 & 2.046 \\
HDF-BX1650 & 12:37:24.11 & $+$62:19:04.7 & 23.24 & 0.18 & 0.74 & 2.100 & 2.093 \\
Q1623-BX470 & 16:25:52.80 & $+$26:43:17.5 & 22.80 & 0.20 & 0.93 & 2.193 & 2.182 \\
Q2343-BX436 & 23:46:09.07 & $+$12:47:56.0 & 23.07 & 0.12 & 0.47 & 2.332 & 2.326 \\
Q1623-BX274 & 16:25:38.20 & $+$26:45:57.1 & 23.23 & 0.25 & 0.89 & 2.415 & 2.405 \\
HDF-BX1485 & 12:37:28.12 & $+$62:14:39.9 & 23.29 & 0.35 & 0.96 & \nodata & 2.551 \\
\enddata
\tablenotetext{a}{Redshift of the Lyman $\alpha$ emission line, when observed.}
\tablenotetext {b}{Average redshift defined by the interstellar absorption lines.}
\end{deluxetable*}

Excluding QSOs which were already known in our survey fields 
(these QSOs are all brighter than ${\cal R}\simeq 20$),
we have identified 21 QSOs and seven narrow-lined AGN 
among the sample of 863 objects with $z > 1$.
This AGN fraction of 3.2\% is essentially the same
as that deduced by Steidel et al. (2002) for LBGs at $z \simeq 3$.
However, we caution that we have not yet quantified the
AGN selection function imposed by our color criteria and spectroscopic
follow-up, so that the above estimate is only indicative at this stage.
On the other hand, it is encouraging that
the redshift distribution of the AGN (both BX and BM)
is similar to that of the galaxies: $\langle z_{\rm AGN} \rangle = 2.31 \pm 0.48$ 
and $\langle z_{\rm GAL} \rangle = 2.14 \pm 0.37$), perhaps suggesting
that differential selection effects may not be too severe.

\subsection{Near-Infrared Spectroscopy}

In parallel with the LRIS-B optical spectroscopy, we have been observing
a subset of the BX and BM galaxies in the near-IR, targeting in particular 
the \ha\ and [\ion{N}{2}]~$\lambda\lambda 6548,6583$ emission lines
in the $H$ and $K$-band. Initial results from this aspect of the survey
were reported by Erb et al. (2003). 

Briefly, we use the
near-infrared echelle spectrograph (NIRSPEC) on the 
Keck~II telescope (McLean et al. 1998) with a
0.76\arcsec$\times$42\arcsec\ entrance slit
and medium-resolution mode. 
In the $K$-band, this set-up
records a $\sim 0.4 \mu$m wide portion of the 
near-IR spectrum at a
dispersion of 4.2\,\AA\ per $27 \mu$m pixel;
the spectral resolution is $\sim 15$\,\AA\ FWHM
(measured from the widths of emission lines from the night sky).
In the $H$-band, $\sim 0.29 \mu$m are recorded 
at a dispersion of 2.8\,\AA\ per pixel and
$\sim 10$\,\AA\ resolution.
The 42\arcsec\ length of the rotatable slit is normally
sufficient to include two galaxies
by choosing the appropriate position angle
on the sky. Individual exposures are 900\,s and we typically
take between two and four exposures per galaxy (or pair of galaxies),
offsetting the targets along the slit by a few arcseconds
between exposures.
The data are reduced and calibrated
with the procedures described by Erb et al. (2003).

This near-IR spectroscopy has several goals, most of which are complementary
to those achievable with the optical spectra described above:
(1) Determine the redshifts of the nebular emission lines which,
to our knowledge, give the closest approximation to the systemic
redshifts of the galaxies. This is particularly important for
examining the connection between the galaxies and the absorption
lines seen along nearby QSO sight-lines, since uncertainties in the systemic
galaxy redshifts are reduced from $\sim 150-200$ \kms (see Adelberger \et 2003) to $\sim 20-30$ \kms;
(2) Measure the velocity dispersion of the ionized gas and
look for evidence of ordered motions, such as rotation, in order to estimate the galaxies' dynamical
masses (see Erb \et 2003);
(3) Determine the metallicity of the \ion{H}{2} regions
from consideration of the [\ion{N}{2}]/\ha\ ratio;
(4) Compare the star formation rate deduced from the \ha\
luminosity with that from the far-UV continuum.
Initial results relevant to these topics were 
presented by Erb et al. (2003); in \S4 we briefly touch
on some of them again. More extensive results from the near-IR
spectroscopic follow-up will be presented elsewhere.  

\section{THE SPECTRAL PROPERTIES OF UV-SELECTED GALAXIES AT $1.4 < z < 2.5$}

In this section we illustrate the astrophysical information that can be
deduced from the analysis of the rest-frame far-UV spectra of galaxies
in the `redshift desert', using Q1307-BM1163 (Fig.~3) as an example.
This is admittedly one of the brightest galaxies discovered
in the survey to date (${\cal R} = 21.66$), and the spectrum 
reproduced in Fig.~3 is of higher signal-to-noise ratio than most.
However, it must be borne in mind that this spectrum was recorded 
with only three 1800\,s exposures. Data of similar quality
can be secured for large numbers of galaxies in the sample with integrations
of $\sim 10$ hours, and such observations are well underway.
The character of the spectrum
of Q1307-BM1163 is similar to those of many other BX and BM objects;
thus we expect that our conclusions from its analysis should be applicable
to at least a subset of the galaxies at $1.4 < z < 2.5$, particularly
those at the bright end of the luminosity function.

\subsection{Integrated Stellar Spectra}

In the far-UV spectra of star forming galaxies we see the integrated
light of young stars of spectral types O and B; such spectra are
most effectively analyzed with population synthesis models such as
{\it Starburst99\/}\footnote{Available from http://www.stsci.edu/science/starburst99/}
(Leitherer et al. 1999) which generally provide a good match
to the observed spectral characteristics at high (e.g. Pettini et al. 2000; 
de Mello, Leitherer, \& Heckman 2000), as well as low (Leitherer 2002),
redshifts. This is also the case for Q1307-BM1163.
In Fig.~5 we compare our LRIS-B spectrum of this galaxy 
with that calculated by {\it Starburst99\/} for a continuous star
formation episode which has been on-going for 100 million years,
with solar metallicity and a power law IMF with
Salpeter (1955) slope $\alpha = 2.35$.
The spectrum of Q1307-BM1163 has been reduced to the rest frame
at redshift $z_{\rm H II} = 1.411$ (see \S4.4) and normalized
to the continuum; continuum windows were selected with reference to
the {\it Starburst99\/} model spectrum.

\begin{figure}[htb]
\centerline{\epsfxsize=9cm\epsffile{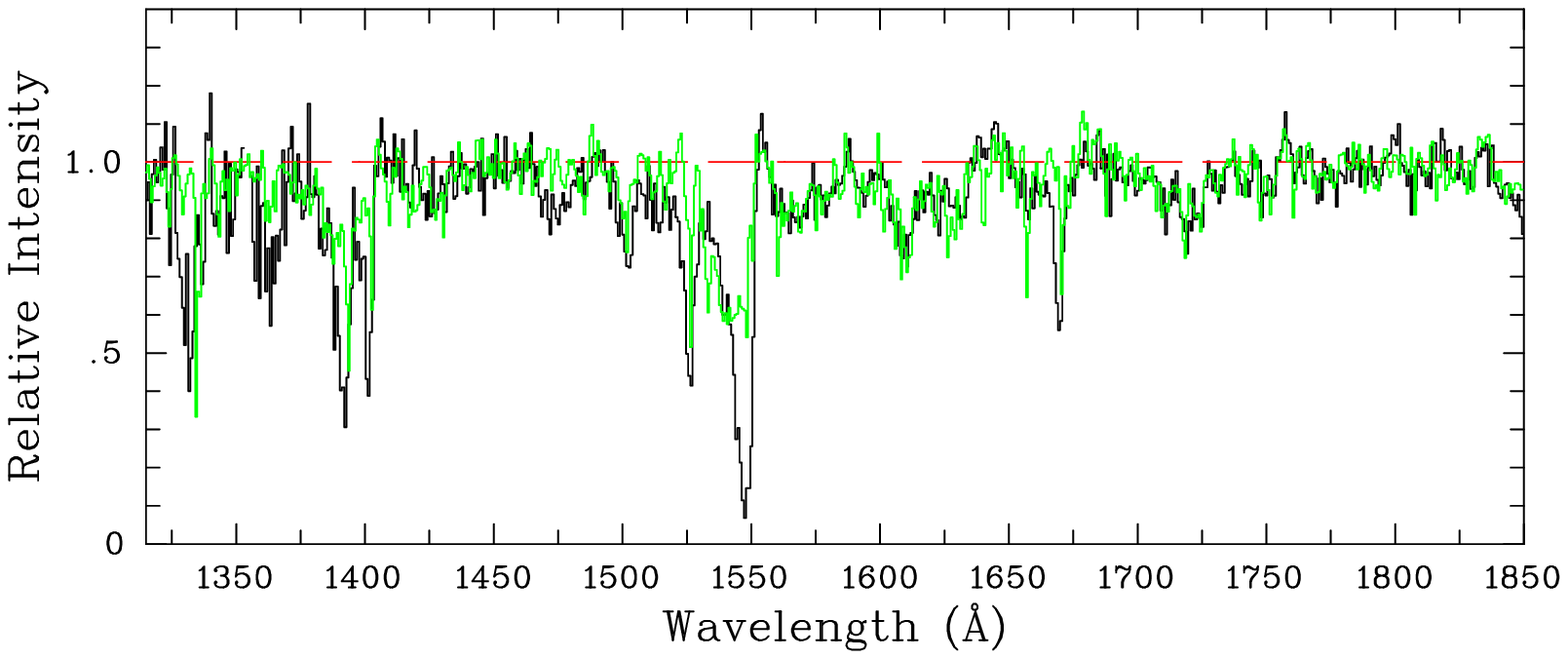}}
\figcaption[f5.eps]{Comparison between the observed far-UV spectrum of Q1307-BM1163
(black histogram),
normalized to the continuum and reduced to rest-frame wavelengths,
and the simplest model spectrum produced with the population synthesis code
{\it Starburst99\/}, one which assumes
continuous star formation, solar metallicity, and a Salpeter
IMF (green or gray histogram). 
The model's match to the data is remarkably good---the 
features which are not reproduced in the {\it Starburst99\/} 
composite stellar spectrum are in most cases
interstellar absorption lines.
}
\end{figure}

This simplest of models---100 million years is the age beyond which
the far-UV model spectrum no longer changes with time---is 
an excellent match to the integrated stellar spectrum of Q1307-BM1163.
The low-contrast blends of photospheric lines are reproduced 
very well, and most of the features which differ between the two
spectra in Fig.~5 are interstellar absorption lines.
{\it Starburst99\/} makes no attempt to reproduce interstellar
features; it is often the case that these lines are stronger 
in the observed spectra of star forming galaxies than
in the models because the latter use empirical libraries of stellar
spectra of relatively nearby OB stars, whereas the observations
sample much longer pathlengths through a whole galaxy.

\begin{figure}[htb]
\centerline{\epsfxsize=9cm\epsffile{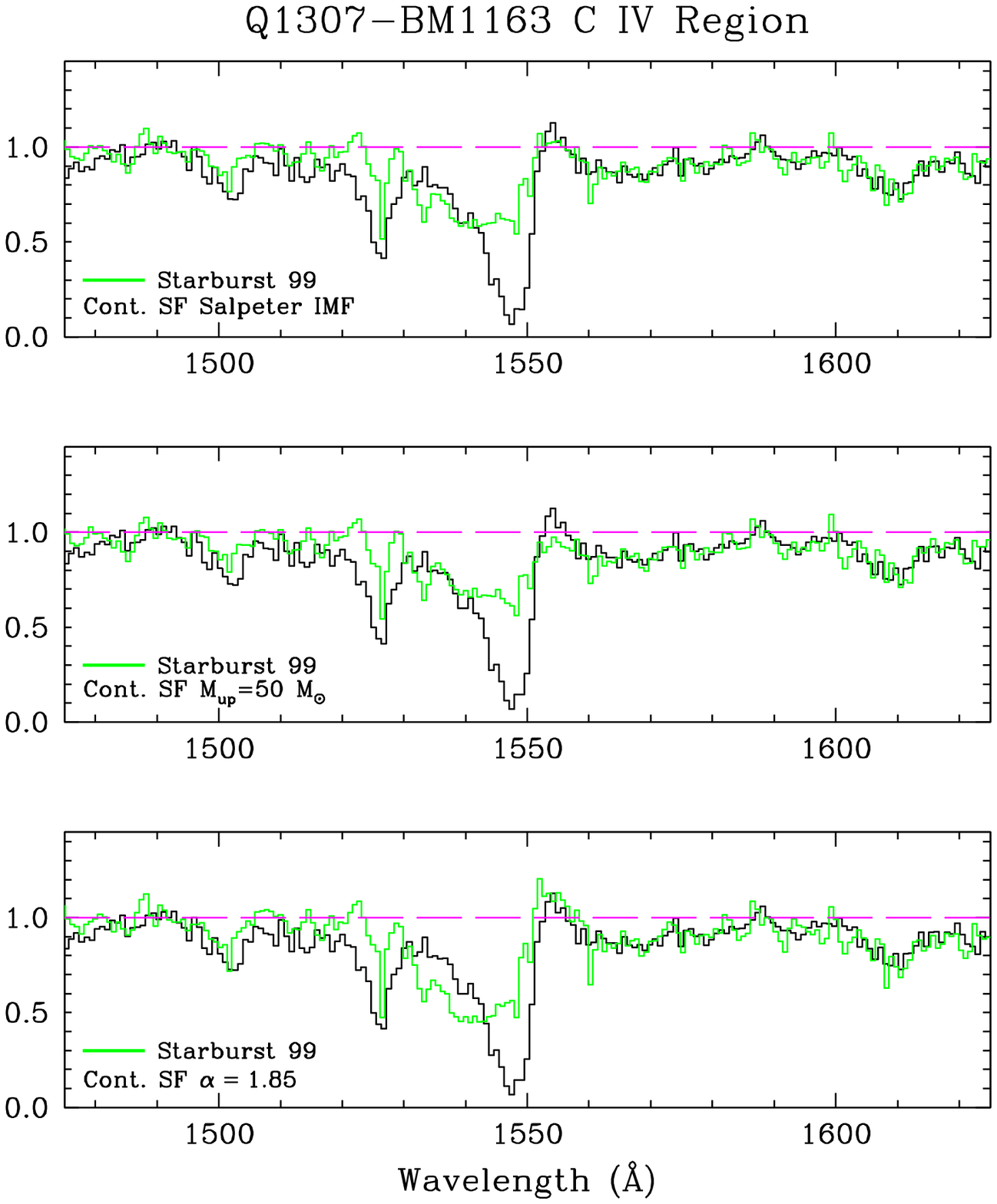}}
\figcaption[f6.eps]{Sensitivity of the \ion{C}{4} P-Cygni profile
to the upper end of the IMF. Black histogram: portion of the
observed spectrum of Q1307-BM1163; green (or gray) histogram:
model spectra produced by {\it Starburst99\/} with, respectively,
a standard Salpeter IMF (top panel), an IMF lacking stars more massive
than $50 M_{\odot}$ (middle panel), and an IMF flatter than 
Salpeter (bottom panel). These changes were deliberately 
chosen to be relatively small, to illustrate the fact 
that it is possible to discriminate between them on the basis
of even a relatively short exposure spectrum of this galaxy.
Overall, there is no evidence for a departure from
the standard Salpeter IMF at the upper end of the mass distribution
of stars in Q1307-BM1163.
}
\end{figure}

Figure~6 shows the comparison on an expanded scale, centered on the
\ion{C}{4}$\lambda\lambda1548,1550$ doublet which is a blend
of P-Cygni emission/absorption from the expanding atmospheres
of massive stars, and narrower interstellar absorption.
The dependence of mass-loss rate on both stellar luminosity and 
the ionization parameter that is modulated by stellar temperature 
makes the \ion{C}{4} P-Cygni profile most pronounced 
in the most massive stars (e.g. Walborn, Nichols-Bohlin, \& Panek 1985).
Thus, its strength relative to the underlying continuum light,
produced collectively by all of the O and early B type stars,
is very sensitive to the slope and upper end cut-off of the IMF,
as illustrated in Fig.~6. Even relatively minor 
changes in $\alpha$ and $M_{\rm up}$ alter the
appearance of the P-Cygni profile: excluding stars with
masses $M > 50 M_{\odot}$ reduces both emission 
and absorption components (middle panel of Fig.~6),
while increasing the proportion of stars at the upper end
of the IMF, by changing $\alpha$ from 2.35 to 1.85, 
over-produces the P-Cygni feature (bottom panel of Fig.~6).
The best overall agreement (from among these three models) is obtained with a standard Salpeter IMF
(top panel).

To a lesser degree, the relative strength of 
the \ion{C}{4} P-Cygni profile
is also sensitive to other parameters,
in particular to the metallicity (Leitherer et al. 2001), since
metal-poor stars experience lower mass-loss rates, 
and to age-dependent dust extinction (Leitherer, Calzetti, \& Martins 2002).
While one could imagine contrived scenarios in which all of these
effects (IMF, metallicity, and dust obscuration)
somehow balance each other, the most straightforward
conclusion from the comparisons in Fig.~6 is that
the metallicity of the early-type stars in 
Q1307-BM1163 is close to solar, and that the youngest stars
do not suffer, overall, significantly higher extinction 
than the whole OB population.
Evidently, at a redshift $z = 1.411$, which corresponds to a look-back time
of $\sim 10$\,Gyr, Q1307-BM1163 had already evolved to a stage
where its young stellar population closely resembled  
the Population~I stars of the Milky Way, at least in their
spectral characteristics.

\subsection{Metallicity}

\subsubsection{Stellar Abundances}

While we have concluded that the metallicity of Q1307-BM1163 is likely
to be close to solar, the wind lines are not ideal for
abundance measurements because they respond to several other parameters,
as explained above.
On the other hand, the far-UV spectrum of star forming galaxies
is so rich in stellar photospheric lines, as can be readily
appreciated from Fig.~5, that it is worthwhile considering
whether any of these can be used as abundance indicators.
Since all of these features are blends of different
lines, this question is best addressed with spectral synthesis
techniques. For example, Leitherer et al. (2001) have used
{\it Starburst99\/} to show that
the blend of \ion{Si}{3}~$\lambda 1417$, \ion{C}{3}~$\lambda 1427$,
and \ion{Fe}{5}~$\lambda 1430$, which they define as the `1425'
index, becomes stable after $\sim 50$\,Myr
(that is, its strength no longer depends on age in a continuous
star formation episode) and its equivalent width decreases
by a factor of $\sim 3-4$, from $W_{1425} \simeq 1.5$\,\AA\ to $\sim 0.4$\,\AA,
as the metallicity of the stars drops from Milky Way to Magellanic
Cloud values, that is from $\sim $solar to $\sim 1/4$ solar.
In Q1307-BM1163 we measure $W_{1425} \simeq 1.2$\,\AA\ [adopting the same
continuum normalization as Leitherer et al. (2001)], which suggests 
a near-solar metallicity.

Very recently, Rix et al. (in preparation) have explored the possibility
of extending this type of approach to other photospheric blends
and to a wider range of metallicities.  These authors have identified
a blend of \ion{Fe}{3} lines between 1935\,\AA\ and 2020\,\AA, and 
defined a corresponding `1978' index which is
potentially very useful for two reasons. 
First, it is considerably stronger 
than `1425' index of Leitherer et al. (2001)
and can thus be followed 
to lower metallicities. Specifically, at metallicity
$Z = Z_{\odot}$, $W_{1978} \simeq 5.9$\,\AA, or 
$\sim 4 \times W_{1425}$, and even at
the lowest metallicity considered by Rix et al.,
$Z = 1/20 Z_{\odot}$, $W_{1978} \simeq 2$\,\AA, which is
greater than $W_{1425}$ at $Z = Z_{\odot}$.
Second, this index is in a `clean' region of the spectrum,
where there are no strong interstellar, nor stellar wind, lines 
to complicate its measurement. In Q1307-BM1163 we measure
$W_{1978} \simeq 6.2$\,\AA\, which again implies 
that the metallicity of the young stars is close to solar.

The measurement of stellar abundances from UV spectral indexes
is a technique which is still very much under development.
It would be premature, for example, to use the 
above values of $W_{1425}$ and $W_{1978}$ 
to draw conclusions concerning the relative
abundances of different elements. 
Nevertheless, these initial indications 
are certainly promising and raise the possibility that, 
once the indices are calibrated with
independent abundance measures,
it may be possible to determine stellar abundances 
to within a factor of $\sim 2$ for large numbers 
of high redshift galaxies from their rest-frame UV
spectra, even if of 
only moderate signal-to-noise ratios.

\subsubsection{Interstellar Gas Abundances}

The most obvious way to calibrate the UV spectral indices
is with reference to nebular abundances determined
from the familiar rest-frame optical emission lines
from \ion{H}{2} regions. While nebular diagnostics
generally apply to different elements (mostly oxygen),
to a first approximation they should give the same
`metallicity' as the young stars which have recently formed 
out of the interstellar gas seen in emission.
This is one of the scientific motivations to obtain 
near-IR spectra of a subsample of BX and BM galaxies
(\S4.3).

\begin{figure}[htb]
\centerline{\epsfxsize=9cm\epsffile{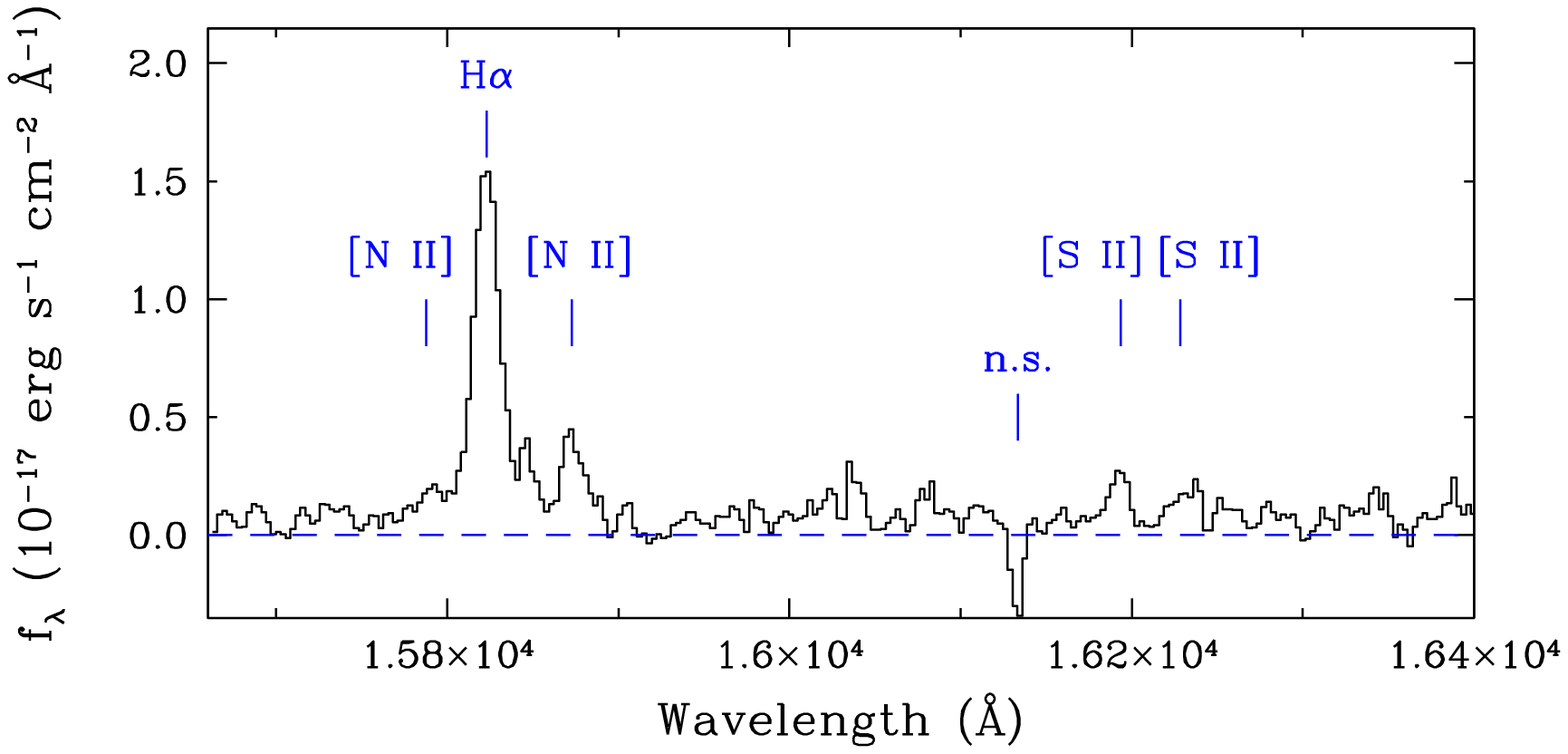}}
\figcaption[f7.eps]{Portion of the NIRSPEC $H$-band spectrum of Q1307-BM1163,
obtained with a 1800\,s exposure through the 0.57\,arcsec slit.
Note that we detect the rest-frame optical continuum from
the galaxy, as well as the emission lines indicated. 
The feature marked `n.s.' is a residual
from the subtraction of a strong sky line.
}
\end{figure}

Figure~7 shows a portion of the $H$-band spectrum of
Q1307-BM1163 encompassing the H$\alpha$,
[\ion{N}{2}]~$\lambda\lambda 6548, 6583$,
and [\ion{S}{2}]~$\lambda \lambda 6716, 6731$ emission lines.
After subtracting the faint continuum,
and fitting H$\alpha$ and [\ion{N}{2}]~$\lambda 6583$
with Gaussian profiles, we measure
a flux ratio [\ion{N}{2}]/H$\alpha = 0.22 \pm 0.06$
($1 \sigma$ random error).
Denicol\'{o}, Terlevich \& Terlevich (2002) have shown that,
in a statistical sense, this ratio scales with the
oxygen abundance. Adopting the recent calibration
by Pettini \& Pagel (2004):

\begin{equation}
{\rm 12+log(O/H)}~=~8.90 + 0.57 \times N2
\end{equation}

\noindent where $N2 \equiv \log ({\rm [N~II]}~\lambda 6583/$H$\alpha$),
we deduce ${\rm 12+log(O/H)} = 8.53 \pm 0.25$
[the error includes both the random error in our
measurement of the $N2$ index and the
$\pm 0.2$\,dex accuracy of the $N2$ calibrator
(at the 68\% confidence level)].
This value of (O/H) is consistent, within the errors,
with the the most recent estimates of the
abundance of oxygen in the Sun,
${\rm 12+log(O/H)} = 8.66 \pm 0.05$
(Allende-Prieto, Lambert, \& Asplund 2001; Asplund et al. 2003),
and in the Orion nebula, ${\rm 12+log(O/H)} = 8.64 \pm 0.06$
(Esteban et al. 1998; 2002).
It would also be of interest to measure element abundances 
in the cool interstellar medium which produces the numerous
absorption lines indicated in Fig.~3. This, however, will
require data of higher spectral resolution, giving access
to weaker absorption lines which are not saturated.
In any case, such an analysis would only yield relative,
rather than absolute, abundances because the Ly$\alpha$
absorption line falls at 2931\,\AA, below the atmospheric cut-off.

Summarizing the results of \S5.2, we have estimated the metallicity of
Q1307-BM1163 from stellar wind lines, stellar photospheric absorption lines,
and emission lines from ionized gas, and all these 
different indicators concur in pointing to an approximately solar
abundance. Preliminary results from our on-going near-IR spectroscopic survey
indicate that galaxies with $Z \simeq Z_{\odot}$
are not unusual at redshifts $z = 1.4 - 2.5$, at least among
the brighter BX and BM objects. 
In Q1307-BM1163 we seem to have an example of a galaxy which 
had already reached solar metallicity $\sim 10$\,Gyr
ago while continuing to support an extremely vigorous star formation,
at a rate of more than $\sim 30 M_{\odot}$~yr$^{-1}$ (see \S5.3). 
This mode of star formation is very different
from that undergone by 
the Milky Way disk at any time in its past
(Freeman \& Bland-Hawthorn 2002)
and it is likely that the descendents of objects like
Q1307-BM1163 are to be found among today's elliptical 
galaxies and bulges of massive spirals.

\subsection{Star Formation Rate}

From the spectrum reproduced in Fig.~7, we measure an 
\ha\ flux $F_{\rm H\alpha} = (2.9 \pm 0.1) \times 10^{-16}$\,erg~s$^{-1}$~cm$^{-2}$.
In the adopted cosmology, this implies an 
\ha\ luminosity $L_{\rm H\alpha} = (3.5 \pm 0.1) \times 10^{42} h^{-2}_{70}$\,erg~s$^{-1}$.
With Kennicutt's (1998) calibration
\begin{equation}
\textup{SFR } (\msun \textup{ yr}^{-1}) = 7.9 \times 10^{-42} \;L_{\Ha} \:\:(\textup{erg s}^{-1})
\end{equation}
we then deduce a star formation rate SFR\,$= (28 \pm 1) M_{\odot}$~yr$^{-1}$.

An independent estimate of the star formation rate is provided
by the UV continuum at 1500\,\AA
\begin{equation}
\textup{SFR } (\msun \textup{ yr}^{-1}) = 1.4 \times 10^{-28}\;L_{1500} \:\:(\textup{erg s}^{-1}\; \textup{Hz}^{-1})
\end{equation}
(Kennicutt 1998); both eq. (4) and (5) assume continuous star formation 
with a Salpeter slope for the IMF from 0.1 to 100\,$M_{\odot}$.
At $z = 1.411$, 1500\,\AA\ corresponds to an observed wavelength
of 3617\,\AA, close to the center of the bandpass of our 
$U_n$ filter (Steidel et al. 2003). Then, the measured
$U_n = 22.22$ (AB) of Q1397-BM1163 corresponds to
SFR\,$= 30 M_{\odot}$~yr$^{-1}$.
The good agreement between the values of SFR
deduced from the \ha\ and the far-UV continuum luminosities 
is not unusual for bright UV-selected galaxies (Erb \et 2003).
To a first approximation, it 
presumably indicates that the UV continuum does not
suffer a large amount of extinction by dust (see the discussion
of this point by Erb et al. 2003).

\subsection{Kinematics of the Interstellar Medium}

From the Gaussian fit of the H$\alpha$ line in Q1307-BM1163
we deduce a one-dimensional velocity dispersion of the 
ionized gas $\sigma = 126$\,km~s$^{-1}$ and a redshift
$z_{\rm H II} = 1.4105$. The former is close to the mean
$\langle \sigma \rangle = 110$\,km~s$^{-1}$
of the sample of 16 (mostly BX) galaxies at 
$\langle z_{\rm H II} \rangle = 2.28$ analyzed by
Erb et al. (2003). It is however significantly higher 
than the mean $\langle \sigma \rangle = 78$\,km~s$^{-1}$
of the 16 Lyman break galaxies at $z \simeq 3$
studied by Pettini et al. (2001), and in fact exceeds
the highest value found in that sample, 
$\sigma = 116 \pm 8$\,km~s$^{-1}$.
Erb et al. (2003) commented on the apparent increase
in velocity dispersion of star forming galaxies 
between $z \sim 3$ and $\sim 2$; it will be interesting
to explore such kinematic evolution in more detail and
to lower redshifts once our near-IR survey of BX and BM 
galaxies is more advanced.

Another important aspect of the internal kinematics 
of LBGs are the large velocity differences
which are nearly always  measured between 
interstellar absorption lines, nebular emission lines,
and \lya\ emission.
If we take the nebular lines to be at the
systemic redshifts of the galaxies, 
the interstellar absorption lines 
and \lya\ are respectively blue- and red-shifted by
several hundred km~s$^{-1}$ 
(Pettini et al. 2001; Shapley et al 2003). 
In this respect also Q1307-BM1163 is no exception-- the numerous interstellar lines in
the spectrum have centroids that are blue-shifted by 300 \kms with respect to the
redshift defined by the H$\alpha$ emission line, and have velocity widths of $\sim 650$ \kms; 
both of these values are quite typical [cf. Pettini \et 2001, 2002, Shapley \et 2003].)
This kinematic pattern 
is most simply explained as being due to large-scale
outflows from the galaxies, presumably powered by
the energy deposited into the ISM by the star formation
activity. The resulting `superwinds' 
are likely to have a far-reaching impact on the
surrounding intergalactic medium and are probably
at the root of the strong correlation 
between LBGs and IGM metals found
by Adelberger et al. (2003).

Indeed, a major motivation for pursuing galaxies in the
`redshift desert' is to investigate how the galaxy-IGM connection
evolves from $z \sim 3$ to lower redshifts. 
We can already address one aspect of this question by examining
the velocity differences between interstellar absorption,
nebular emission, and \lya\ in 27 BX and BM galaxies 
which we have observed at \ha\ with NIRSPEC and which also
have LRIS-B spectra of sufficiently high quality to measure
absorption and (when present) \lya\ redshifts with confidence.

\begin{figure}[htb]
\centerline{\epsfxsize=9cm\epsffile{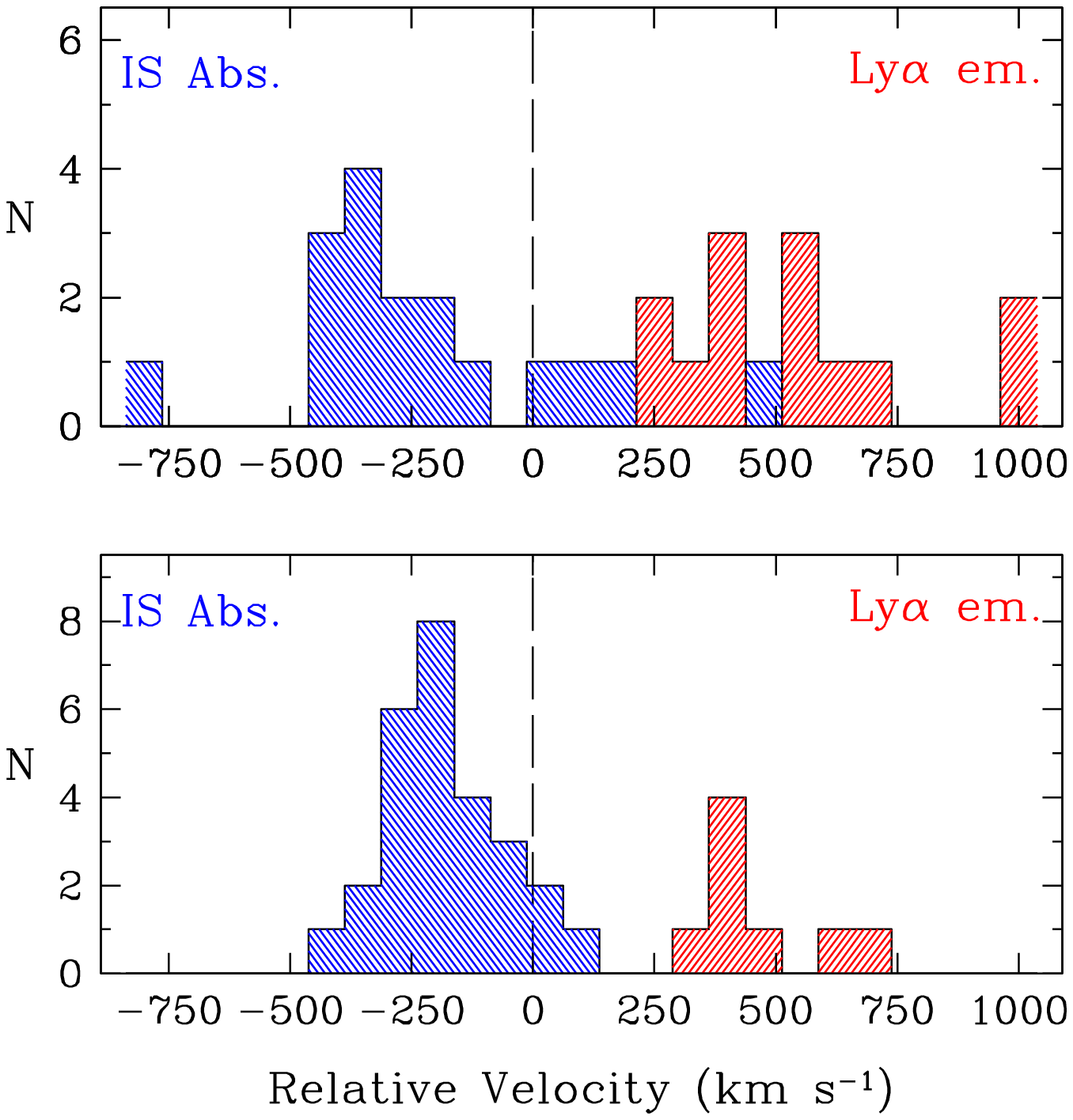}}
\figcaption[f8.eps]{Velocity offsets of the interstellar absorption lines
(blue-hatched histogram) and, when present, \lya\ emission (red-hatched histogram)
relative to the systemic redshifts defined by the nebular emission lines
(vertical long-dash line).
The top panel shows the results for $z\sim3$ LBGs presented by Pettini et al.(2001); 
the bottom panel shows results for a sub-sample of 27 BX and BM galaxies
which have been observed with NIRSPEC at \ha\ and also have high quality 
LRIS-B spectra. The velocity offsets seen in BX and BM galaxies are
similar, in both magnitude and distribution, to those typical
of LBGs at $z \sim 3$.
}
\end{figure}

The results of this exercise are shown in Fig.~8.
For these 27 galaxies, the mean velocity offset of the
interstellar lines is 
$\langle \Delta v_{\rm IS, abs} \rangle = -175 \pm 25$\,km~s$^{-1}$.
For the ten galaxies among them which exhibit detectable
\lya\ emission, $\langle \Delta v_{\rm Ly\alpha} \rangle = +470 \pm 40$\,km~s$^{-1}$.
These values, and their observed
distributions, are very similar to those found at $z \sim 3$ 
by Pettini et al. (2001) and Shapley et al. (2003)---a comparison
with the data presented in Pettini et al. (2001) is included in Fig. 8.
Thus, the superwinds generated in active sites of star formation 
appear to have similar kinematic characteristics from $z \sim 3$
down to at least $z \sim 1.5$, even though their parent galaxies 
may become more massive over this redshift interval, if the hints
provided by the nebular line widths have been correctly interpreted
(Erb et al. 2003).
These tentative conclusions
make it all the more interesting 
to investigate how the 
galaxy-IGM connection may evolve to lower redshifts.

\section{NEAR-IR PHOTOMETRIC PROPERTIES}

We end with a brief comment on the  
$K_s$-band magnitudes and ${\cal R}-K_s$
colors of UV-selected star-forming galaxies in the `redshift desert'.
In 2003 June we initiated a program of deep
$K_s$-band photometry of these galaxies using
the Wide Field Infrared Camera 
(WIRC) on the Palomar 5.1\,m Hale telescope.
The camera employs a
$2048 \times 2048$ Rockwell HgCdTe array, and has a field of view
of 8\minpoint7 $\times$ 8\minpoint7 
with a spatial sampling of 0\secpoint25 per pixel. 
With $\sim 12$ hour integrations,
the images reach 5$\sigma$ photometric limits
(in 2 arcsec diameter apertures) of $K_s \simeq 21.7$, 
sufficiently deep to detect $\sim 85$\% 
of the galaxies with spectroscopic redshifts.
In Fig. 9 we show initial results from the first three WIRC
pointings (in the Q1623+27, Q1700+64, and Q2343+12 fields,
where only BX candidates have so far been observed spectroscopically);
while preliminary, these data already allow a coarse comparison with 
other faint galaxy samples.

\begin{figure*}[htb]
\centerline{\epsfxsize=6cm\epsffile{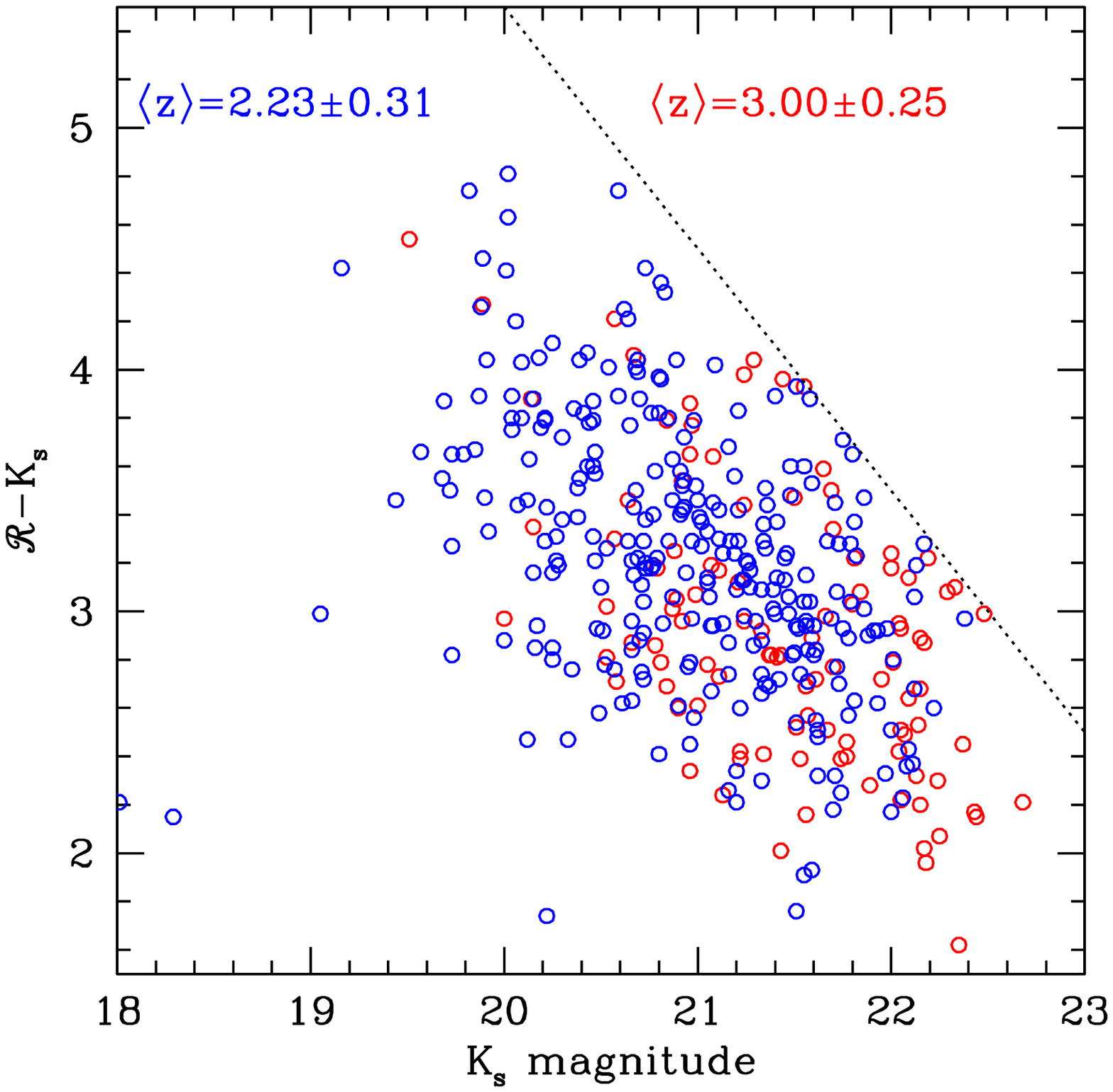}\epsfxsize=6cm\epsffile{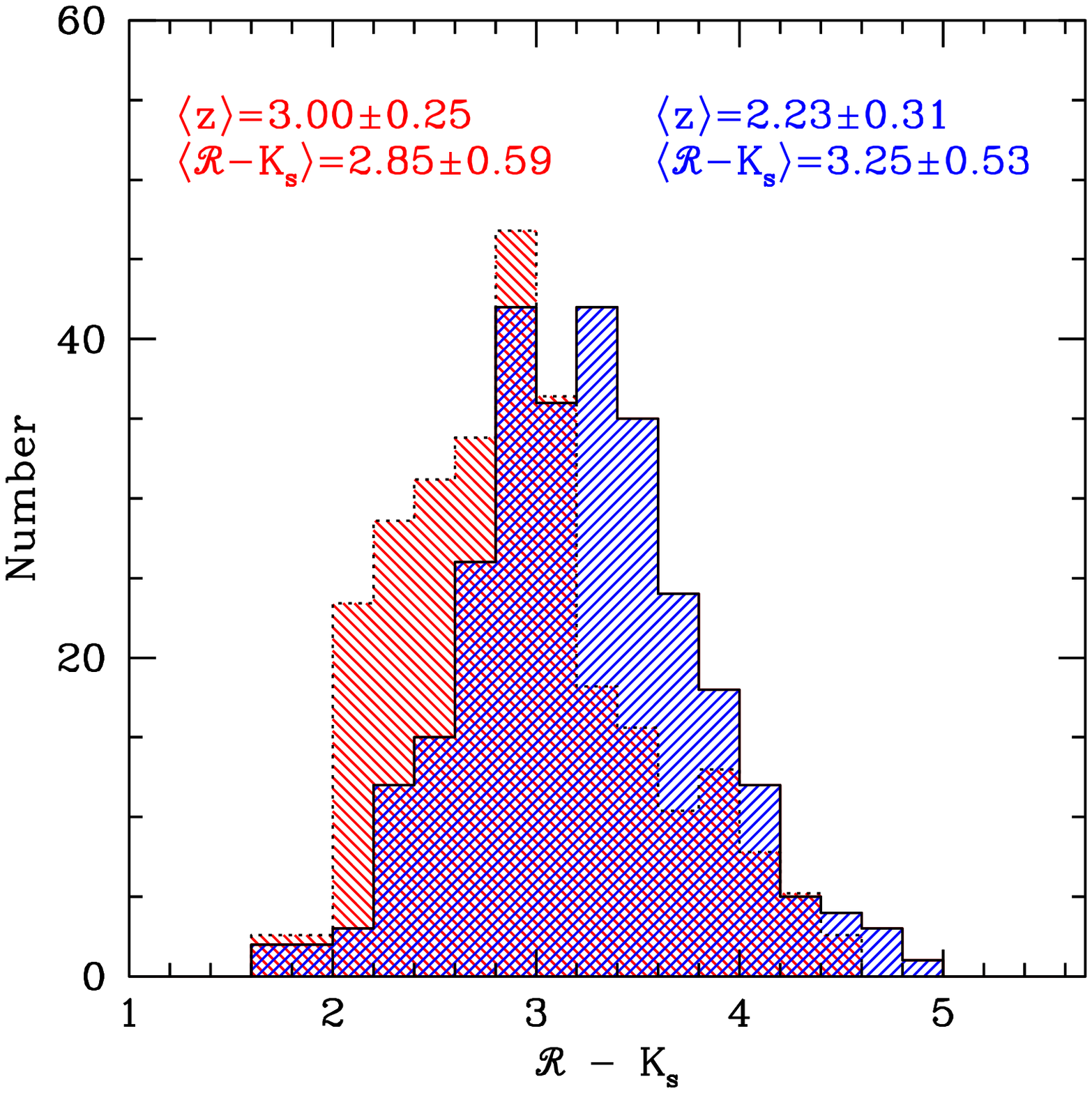}}
\figcaption[]{{\it Left Panel:\/}~ 
Color-magnitude diagram for the 283 BX galaxies with $K_s$ band 
measurements to date from the Q1623, Q1700, and Q2343 fields (blue points);
the 5$\sigma$ detection limit is $K_s \simeq 22$. The red points
are the sample of 108 galaxies at $z \sim 3$ from Shapley \et 2001.  
The dotted line indicates galaxies with ${\cal R}=25.5$, the limit
for both spectroscopic samples. 
Approximately 15\% of the BX galaxies with spectroscopic redshifts
were not significantly detected in the images; most of these have optical
magnitudes ${\cal R}>25.0$. 
{\it Right Panel:\/}~The distributions of ${\cal R}-K_s$ colors for the same 
BX sample at $\langle z \rangle = 2.2$ (blue) as compared to that of
the $z \sim 3$ LBGs studied by Shapley \et (2001). 
The BX galaxies are significantly redder in their optical/IR colors
despite having been selected to span the same range of UV color as the
LBGs (see text for discussion).
}
\end{figure*}



As can be seen from the left-hand panel of Fig. 9, 
$\sim 10$\% of UV-selected galaxies at $z \sim 1.9-2.5$
are brighter than $K =20$;
thus we expect a relatively small overlap between the BX population
and the high redshift tail of published $K$-selected
samples (Cohen et al. 1999ab; Cimatti \et 2002; Daddi et al. 2003).
However, going only one magnitude deeper to $K \sim 21$
should pick up an appreciable fraction of the spectroscopic BX sample. 

It is intriguing to find a clear difference in the
${\cal R}-K_s$ colors of BX galaxies at $z \sim 2.2$
and LBGs at $z \sim 3$ (Shapley \et 2001),
in the sense that the former are significantly redder than the
latter, on average 
(see right-hand panel of Fig.~9). 
For a given star formation history and extinction, galaxies at $z \sim 2.2$ and
$z \sim 3.0$ would have identical ${\cal R}-K_s$ color (the k-corrections are identical in the two
bands for model SEDs that fit the observed colors), meaning that whatever the cause,
there is a significant intrinsic color difference in the UV-selected galaxies in the two redshift intervals. 
Since the $K_s$ and ${\cal R}$ filters straddle the age sensitive
Balmer break at these redshifts,  
one interpretation of the redder colors
would be that, on average, star formation has been proceeding for longer periods of
time in the $z \sim 2$ galaxies as compared to similarly-selected galaxies at $z \sim 3$. If
the $z \sim 3$ LBGs continued to form stars during the $\sim 800$ Myr interval between
the two epochs, such reddening of the ${\cal R}-K_s$ color would 
be expected. 
However, we caution that there may be other reasons
for the offset evident in Fig.~9, possibly related to 
higher dust extinction and/or larger contamination of the broad-band colors
with line emission (most of the $z \sim 2$ galaxies would have H$\alpha$ in the K band, whereas
most of the $z \sim 3$ galaxies had [OIII] and H$\beta$ in the K band--although these tend to have
roughly the same equivalent widths for the sub-samples that have been spectroscopically
observed in the near-IR), 
as well as to the way the different samples were selected\footnote{The $z \sim 3$ sample of
Shapley \et 2001 over-sampled the optically brightest galaxies and galaxies having the reddest
UV colors relative to a random ${\cal R}$-selected spectroscopic sample, so that the two
samples may not be exactly analogous. Possible differential selection effects will be better quantified
in future work.}  
At present, both the observed increase in one-dimensional
line widths and the reddening of the optical/IR colors are qualitatively consistent with 
a significant overall increase in stellar mass among at least a substantial fraction of
the UV-selected
populations between $z \sim 2$ and $z \sim 3$\footnote{Very recently, an increase in the stellar
mass of UV-selected galaxies has been inferred between $z \sim 4$ and $z \sim 3$ in the GOODS-S field, using similar
arguments; Papovich \et 2003}.
Ongoing near-IR spectroscopy targeting BX (and eventually BM)
galaxies with red ${\cal R}-K_s$ colors, and full modeling 
of the observed
optical/IR SEDs using population synthesis, will allow
us to reach firmer conclusions on the cause of the observed
evolution.

\section{SUMMARY}
The main thrust of this paper has been to show that the
redshift interval $1.4 \simlt z \simlt 2.5$, which has
so far been considered hostile to observations, is in
fact ripe for scientific exploration. Galaxies in
what used to be called the `redshift desert'
can in reality be easily identified from their 
broad-band $U_n G {\cal R}$ colors and can be studied very 
effectively  with a combination of ground-based optical 
and near-IR spectroscopy, provided the optical instrumentation has 
high efficiency in the near-UV. It is then possible
to cover most of the rest-frame UV and optical 
spectra of these galaxies, from \lya\ to \ha, and gain
access to a wider range of important spectral diagnostics
than is usually available for the so-far better studied
Lyman break galaxies at $z \sim 3$ (or even galaxies at $z \simlt 1$).
In addition, because the luminosity distances are lower,
the galaxy luminosity function can be probed $\sim 0.6 - 1.1$
magnitudes deeper than is the case at $z \sim 3$, 
and there are more galaxies brighter than 
${\cal R} \sim 23.5$ whose spectra can be recorded at high
spectral resolution and S/N for further, detailed,
investigation.

We have presented the first results from our survey
for galaxies at $1.4 \simlt z \simlt 2.5$, 
in seven fields totaling $\sim 0.5$ square degrees; five of the fields were chosen because
they include one or more bright background QSOs. Over this
area, we have identified thousands of candidates
which satisfy newly defined BX and BM color selection
criteria and have spectroscopically confirmed 863 of them to be at
$z > 1$; 692 are in the targeted $z = 1.4 - 2.5$ range.
The rest-frame UV spectra of BX and BM galaxies 
are very similar to those of LBGs, with a rich
complement of stellar and interstellar lines.
There seem to be proportionally fewer galaxies
with detectable \lya\ emission, but we have not
established yet whether this is related to the 
color selection cuts we have adopted or is a real 
effect. 
The fraction of faint AGN within this sample is 3.2\%,
essentially the same as in the LBG sample at $z \sim 3$.

We have illustrated the range of physical
properties which can be investigated
with the combination of rest-frame UV and optical spectroscopy
using, as an example, one of the brightest objects in the
sample, Q1307-BM1163. This $z = 1.411$ galaxy is 
forming stars at a rate
$\sim 30 M_{\odot}$~yr$^{-1}$ and with a Salpeter 
slope at the upper end of the IMF.
Various abundance indicators,
based on stellar wind and photospheric 
lines, show that the metallicity of the
youngest stars is close to solar;
this is in good agreement, as expected, with
the solar abundance of oxygen in its \ion{H}{2}
regions implied by the high [\ion{N}{2}]/\ha\ 
ratio. We draw attention to the potential 
for abundance determinations of
newly developed UV spectral indices which 
measure the strengths of 
blends of photospheric lines;
once properly calibrated, these indices
may allow the metallicities of large
numbers of galaxies to be approximately assessed from
spectra of only moderate signal-to-noise ratios.
Viewed at a look-back time of $\sim 10$\,Gyr,
Q1307-BM1163 is clearly turning gas into stars,
and enriching its ISM with their products,
at a much faster rate than that experienced by the
Milky Way disk at any time in its past.
We speculate that by $z = 0$ it will have become
an elliptical galaxy or perhaps the bulge of a
massive spiral.

The galactic-scale winds which are commonly 
seen in LBGs at $z \sim 3$ are still present
in star-forming galaxies at later epochs,
generating velocity differences of several
hundred km~s$^{-1}$ between absorption lines
produced by the outflowing ISM and the emission
lines from the star-forming regions.
The typical velocity difference of $\sim 200-300$\,km~s$^{-1}$
between emission and absorption does not seem
to change between $z \sim 3$ and $\sim 2$,
even though there is evidence that the velocity
dispersion of the ionized gas increases by
$\sim 40$\% between these two epochs, possibly
reflecting a growth in the typical galaxy mass.

Initial results from deep $K_s$-band imaging of
spectroscopically confirmed BX galaxies show that
a larger proportion of the $z \sim 2$ galaxies have
relatively red ${\cal R}-K_s$ colors as would be expected if
the $z \sim 2$ galaxies have been forming stars at close to their
observed rate for a longer period of time than their $z \sim 3$
counterparts (and hence would have correspondingly larger stellar
masses). Approximately 10\% of the BX spectroscopic sample 
are very bright in the near-IR ($K_s < 20$) and thus would be 
expected to comprise part of the high-redshift tail of current $K_s$-selected 
spectroscopic surveys.   

There are several other issues of interest
which can be addressed with a large sample 
of galaxies at these intermediate redshifts,
such as their luminosity function and integrated
star formation rate density, the impact of the
`superwinds' on the galaxies' environment and
the intergalactic medium at large, and the relationship
between galaxy morphology and kinematics.
We intend to consider these topics in the future, as our
survey progresses beyond the initial stages
which have been the subject of this paper.\\

We would like to thank the rest of the team responsible for the design, construction, and
commissioning of the LRIS-B instrument and the upgraded CCD camera,
particularly Jim McCarthy, John Cromer, Ernest Croner, Bill Douglas,
Rich Goeden, Hal Petrie, Bob Weber, 
John White, Greg Wirth, Roger Smith, Keith Taylor, and Paola Amico.
We have benefited significantly from software written by 
Drew Phillips, Judy Cohen, Patrick Shopbell, and Todd Small. 
CCS, AES, MPH, and DKE have been supported by grants AST00-70773 and
AST03-07263 from the
US National Science Foundation and by the David and Lucile Packard Foundation.  
NAR has been supported by an NSF Graduate Fellowship.
KLA acknowledges support from the Harvard Society of Fellows.

\appendix

\section{THE LRIS-B SPECTROGRAPH}

The blue channel of the LRIS spectrograph (LRIS-B) was anticipated 
as an upgrade to the LRIS instrument (Oke \et 1995) from
the initial planning stages of the first-light Keck Observatory 
instrumentation suite, and was begun as a project in 1995. 
LRIS-B was installed on the LRIS 
instrument during the summer of 2000, and saw first-light in 2000 September. 
After the installation of LRIS-B, the spectrograph now provides two independent,
optimized imaging spectrograph channels which simultaneously observe the same 5\minpoint5 by 8\minpoint0 field of view 
in two different wavelength ranges through the use of a dichroic beamsplitter. Figure 10 illustrates a section
view of the LRIS instrument that is color-coded to indicate the light paths through the
red and blue channels. The red side of the
instrument maintains identical performance to the original LRIS spectrograph, and (as always) employs 
reflection gratings to disperse red light that is passed by the beamsplitter. The blue channel uses  
UV/blue optimized grisms as dispersers. 
An early technical description of LRIS-B is given by McCarthy \et (1998);
more recent information is available at 
http://www.keck.hawaii.edu/realpublic/inst/lris/lrisb.html\,.

\begin{figure*}[htb]
\centerline{\epsfxsize=13cm\epsffile{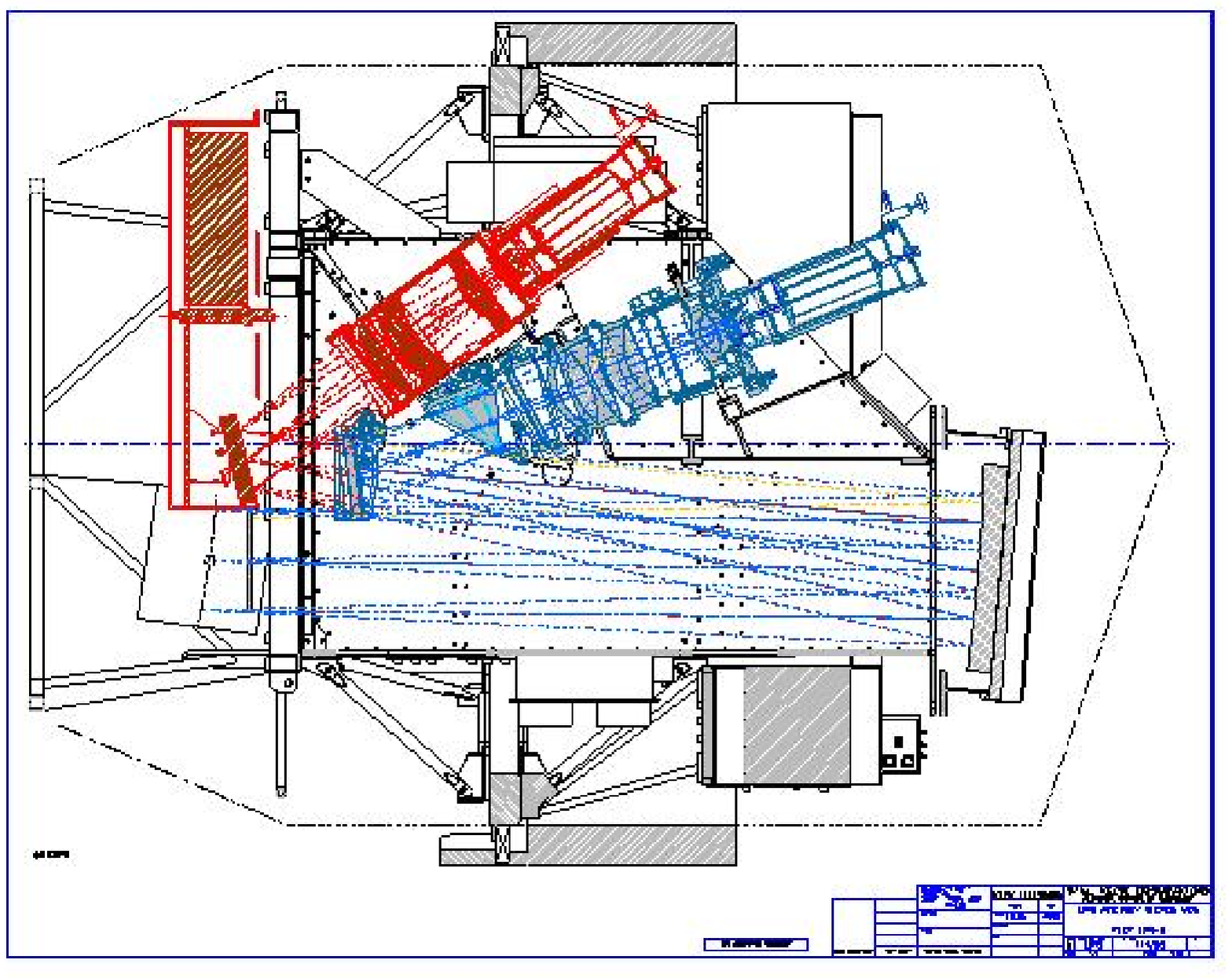}}
\figcaption[f10.eps]{Section view of the LRIS instrument after addition of the LRIS-B components.
The light path is illustrated with incoming rays from the left. The light is collimated
by the mirror (lower right), and then split into blue and red beams at the dichroic. 
The LRIS-B optical elements are shaded in blue; from left to right, they are
the dichroic beam splitter, the grism, the filter, the camera, and the CCD dewar.} 
\end{figure*}

In brief, the LRIS-B upgrade involved the installation
of a UV/blue-optimized spectrograph camera, 
replacement of the camera bulkheads for both the existing spectrograph (LRIS-R) and LRIS-B, 
and installation of independent carousels and transport mechanisms 
to store and deploy blue-side dichroics, grisms, and filters into the beam. 
In addition, all of the LRIS electronics were enclosed inside a glycol-cooled
compartment for better control of the thermal environment at the Cassegrain
focus of Keck I, and the overall instrument software was re-configured 
to accomodate the operations of both blue and red side mechanisms and the simultaneous operation of two 
independent detector trains.  
Each of the optical elements (moving from left to right on figure 10, the dichroic,
grism, and filter) can be changed in less than 60 seconds,
and in particular a grism or filter can be deployed or retracted from the beam
in $\sim 30$ seconds, enabling rapid switching from imaging to spectroscopic
mode which greatly improved the efficiency of slitmask alignment with LRIS.

The instrument currently allows the user to select one of 5 possible dichroic beam-splitting
optics: a flat aluminized mirror, which sends all of the light into the blue camera,
and dichroics that divide the beam at 4600 \AA (d460), 5000 \AA (d500), 5600 \AA (d560),
or 6800 \AA (d680). The reflectivity of the dichroics for wavelengths shortward
of the design cutoff are generally better than 96\% (i.e., superior to the reflectivity
of aluminum), with transmittance better than 90\% longward of the cutoff. 

The LRIS-B filters, which include publicly available u', B, G, and V, are generally
only used for imaging programs or for slitmask alignment images. User-supplied
filters (which must be larger than $\sim 7.5$ by 8 inches in order not to vignette
the parallel beam) may be installed in the filter carousel. With suitable choice of
dichroic and red-side filter, it is possible to image in two passbands
simultaneously. 

There are currently 4 grisms available with LRIS-B: a 1200 line/mm grism blazed at
3400 \AA\ which can cover the wavelength range 3000-4300 \AA with a resolution of
$\sim 3600$ (with a 0.7\arcs\ slit), a 600 line/mm grism blazed at 4000\AA\ ($R \simeq 1600$), 
a 400 line/mm grism blazed at 3400 \AA\ optimized for
the highest throughput at wavlengths $\simlt 4000$\AA\ ($R \sim 1000$), and a 300 line/mm grism 
blazed at 5000\AA\ ($R \sim 900$) . With suitable choice of dichroic beamsplitter and LRIS-red side grating,
it is possible to cover all wavelengths from the atmospheric cutoff near 3000\AA\ to 1 $\mu$m 
with high efficiency.

Because both the red and blue channels of the spectrograph 
share the same paraboloidal reflecting collimator (see figure 10), 
the original protected silver coating on the 0.53m diameter collimator mirror 
was replaced by the hybrid coatings developed by the
group at Lawrence Livermore National Laboratory (Thomas \& Wolfe 2000)  
which provide reflectivity of $\simgt 95\%$ at all wavelengths 
from 3100\,\AA\ to $1\,\mu$m.

The light path for LRIS-B  goes as follows: a slitmask (or long slit) is deployed
in the telescope focal plane, at a position centered 6\arcm\ off axis. After passing
into the intrument and through a field lens, all light is collimated by the 
reflecting collimator.  The dichroic beam splitter (placed just in front of the 
red channel reflection grating) receives the collimated beam and reflects
wavelengths shortward of the dichroic cutoff into the 
blue channel, passing light longward of the cutoff directly onto the red side
grating  (or flat mirror in the case of imaging).
The blue light is then dispersed by a selectable grism, after which it
passes through a selectable filter (tilted by about 6 degrees with respect to the collimated beam
to avoid internal reflections) and into the LRIS-B camera. 
The camera is all refractive, 
constructed from 12 CaFl and fused silica elements with optimized coatings 
that achieve $<$0.5\% reflection
losses at each of the 8 air/glass surfaces.  
The LRIS-B camera has an aspheric first element that is placed slightly
off-axis to correct much of the coma introduced by the parabolic collimator. 
Tests carried out during the commissioning period 
confirmed that the camera achieves images with
RMS image diameters better than 22$\mu$m (0.20\arcsec)  
over the full 8.0\arcmin\ by 5.5\arcmin\ field of view 
and from 3100 to 6000\,\AA\ without the need to re-focus. 
The camera performs well to $\sim 6800$\,\AA, 
beyond which the image quality and throughput of the system deteriorate 
somewhat; in general, this is not a problem since red light is directed
into the red channel of the spectrograph.

\begin{figure}[htb]
\centerline{\epsfxsize=9cm\epsffile{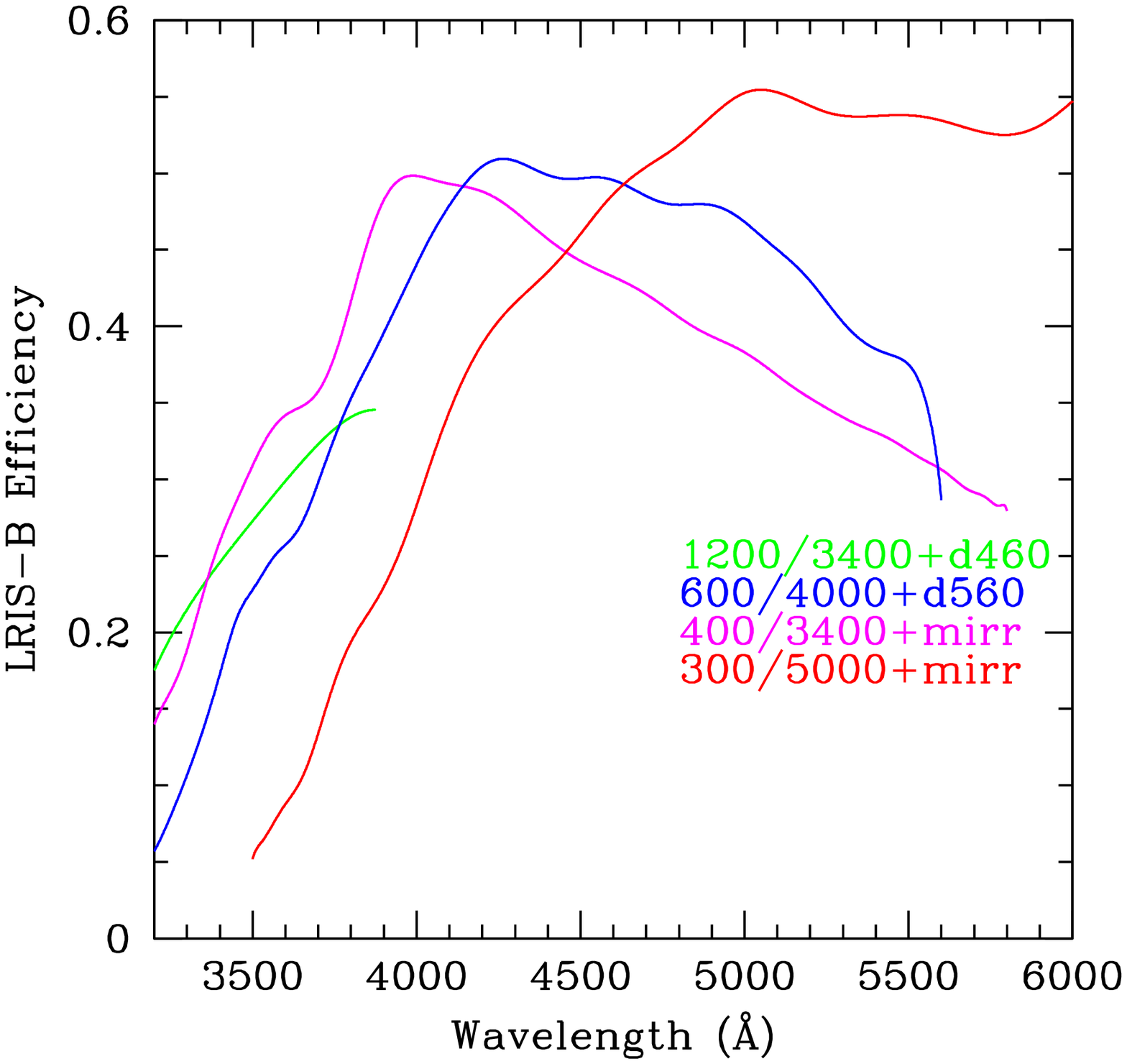}}
\figcaption[f11.eps]{The total spectroscopic system throughput of the LRIS-B instrument
(excluding slit losses and the telescope)
as a function of wavelength for several configurations.
The 400/3400 grism was used in all the observations presented
in this paper because it offers the highest throughput 
between 3100 and 4000\,\AA---the most important wavelength range
for spectroscopic confirmation in the $z=1.4-2.5$ `redshift desert'. Note
that slightly higher UV throughput using this grism is achieved when
a dichroic beam splitter rather than the flat mirror is used.}
\end{figure}

Initially, the LRIS-B detector was an engineering-grade 
SITe $2048 \times 2048$ pixel CCD.
This was replaced in 2002 June
by a science grade mosaic of two EEV (Marconi) 
$2048 \times 4096$ devices selected to have particularly high near-UV 
and blue quantum efficiency. 
With the new detector, LRIS-B records images and spectra over the full field of view
of the instrument with the exception of a 13\arcsec\ inter-chip gap 
that runs parallel to the dispersion direction, and which coincides with the bar 
that is used to support the slit masks in the slit mask
frames. The plate scale at the detector is 0.135\arcsec\ per 15$\mu$m pixel.  
 
To our knowledge LRIS-B is the only UV/blue optimized
faint object spectrograph on an 8-10\,m class telescope. 
It is particularly worthwhile implementing this type of instrument at a site 
such as Mauna Kea which, because of its
high altitude, has atmospheric opacity in the $3100-4000$\,\AA\ 
range that is 20--30\% lower at zenith than that at 
many other observatory sites. Because every optical element has been
optimized for the blue and UV, the instrument achieves remarkably high
efficiency.  The total {\it spectroscopic} system throughput of Keck I+LRIS-B  
measured during the commissioning run with the science-grade CCD mosaic 
averages $\sim 40$\% in the $3800-6000$\,\AA\ range and 
25\% at 3500 \AA; the corresponding values 
for the instrument alone (that is, neglecting telescope losses) 
are higher by about 30\%. The measured instrumental throughput for
a number of grism and dichroic combinations is shown in fig 11.  
LRIS-B represents an increase in spectroscopic efficiency over the original LRIS instrument
(now LRIS-R)
of a factor of $\sim 2.2$ at 4000\AA\, and even at $5000-6000$\AA\ the throughput gain is
$\sim 40\%$ relative to LRIS-R; the gain at wavelengths $\lambda < 4000$\AA\ is more than
a factor of 10 .   
Coupled with the very dark night sky background in the UV-visual
range, LRIS-B allows for unprecedented spectral throughput and is
optimized for very faint objects at intermediate to low spectral
resolution.

At present, there is no atmospheric dispersion corrector (ADC) for the
Keck I Cassegrain, so that the full broad-band capabilities of the
instrument using multislit masks are not yet realized. However, at the
time of this writing an ADC is being designed and contructed at the
University of California Observatories, for deployment in late 2004.

\end{document}